\documentclass[useAMS,usenatbib]{mn2e}

\usepackage{graphicx}

\title{On the tidal dependence of galaxy properties}

\author[Heling Yan, Zuhui Fan And Simon D. M. White]{Heling Yan$^{1}$\thanks{E-mail:
yanheling1984@gmail.com}, Zuhui Fan$^{1}$ and Simon D. M. White$^{2}$
\\
$^{1}$Department of Astronomy, Peking University, YiHeYuan-Road 5, 100871 Beijing, China\\
$^{2}$Max Planck Institute for Astrophysics, Karl-Schwarzschild-Strasse 1, 85740 Garching, Germany}

\begin{document}


\pagerange{\pageref{firstpage}--\pageref{lastpage}} \pubyear{2010}

\maketitle

\label{firstpage}

\begin{abstract}
Using volume-limited samples drawn from The Sloan Digital Sky Survey Data Release 7 (SDSS DR7),
we measure the tidal environment of galaxies, which we characterize by the ellipticity $e$ 
of the potential field calculated from the smoothed spatial number density $1+\delta$ of galaxies.
We analyze if galaxy properties, including color, D$_n$4000, concentration and size correlate with $e$, 
in addition to depending on $1+\delta$. 
We find that there exists a transition smoothing scale at which
correlations/anti-correlations with $e$ reverse.
This transition scale is well represented by the distance to the 3rd-nearest-neighbor of a galaxy in a volume
limited sample with $M_{r} <-20$
which has a distribution peaked at $\sim 2$ $h^{-1}$Mpc. 
We further demonstrate that this scale corresponds to that where the correlation
between the color of galaxies and environmental density $1+\delta$ is the strongest.
For this optimal smoothing $R_{0}$ no additional correlations with $e$ are observed. 
The apparent dependence on tidal ellipticity $e$ at other smoothing scales $R_s$ 
can be viewed as a geometric effect, arising from the cross correlation between $(1+\delta_o)$ and $e(R_s)$. 
We perform the same analysis on numerical simulations with semi-analytical modeling (SAM) of galaxy formation.
The $e$ dependence of the galaxy properties shows similar behavior to that in the SDSS, although the color-density correlation
is significantly stronger in the SAM. The  `optimal adaptive smoothing scale' in the SAM is also closely related to the 
distance to the 3rd-nearest-neighbor of a galaxy, and its characteristic value   
is consistent with, albeit slightly smaller than for SDSS.  
\end{abstract}


\section{Introduction}

While the standard picture of the hierarchical build-up of large-scale structure in the universe has been supported by 
ever increasing observational evidence, 
it is still an ongoing task to understand thoroughly the detailed baryonic physics
associated with galaxy formation, such as gas accretion, star formation and feedback. 
Direct observation of these gaseous processes is challenging.
On the other hand, given the large galaxy samples currently available to us, such as SDSS DR7 \citep{abazajian09},
statistical analysis of galaxy properties has proved to be invaluable to probe the underlying physics affecting galaxy formation and evolution.

The morphology-density correlation for galaxies, uncovered already in the 1930s \citep{hubble36},
shows a clear segregation, with early-type galaxies preferentially 
existing in high density regions
\citep{oemler74,dressler78, postman84,blanton05b,kauffmann04,wienmann06,blanton07,park07}. 
Further investigation has indicated that it is the star-formation related properties, such as 
color and emission-line flux, that are more directly correlated with environmental density 
\citep{kauffmann04,blanton05b,christlein05}. At fixed color, the residual dependence of morphology 
on the environmental density is rather weak \citep{park07,ball08,bamford08}. 
It is also emphasized by, e.g., \citet{kauffmann04}, \citet{blanton07} and \citet{park07}, 
that the environment dependence 
is in fact quite local. While galaxies with abundant close neighbors 
have clearly different properties from those of isolated galaxies, on scales
larger than $\sim 1 h^{-1}\hbox{ Mpc}$, environmental effects probably 
have little influence on galaxy properties \citep{blanton07}.

Theoretically, it has long been known from extensions of Press-Schecter theory that massive halos preferentially reside 
in dense environments \citep{bond91,mo96}. Galaxies therein have properties that are different from those in small halos.
On the other hand, the excursion set theory with sharp-k filter and with dynamics described
by the spherical collapse model predicts that the formation history of halos of fixed mass should
depend only on smaller-scale structure not on their large-scale environment \citep[e.g.,][]{white96}.
Incorporating the ellipsoidal collapse model 
into excursion set theory gives halo mass functions and bias in better agreement with 
numerical simulations \citep{bond96,sheth01}.
In its simple version, however, the theory takes into account the ellipsoidal collapse 
by considering the average tidal field, and thus also predicts no correlations between the formation 
history of halos and their specific large-scale environment \citep{sheth01}. 

Recent numerical simulations show that halo properties can be strongly correlated with their environment. 
Halo assembly bias reveals that old halos tend to cluster more strongly than their younger counterparts, and
this bias is particularly significant for low mass halos \citep{gao05,harker06,wechsler06,gao07,croton07,jing07}.
Using filters other than the sharp-k filter in the excursion set theory leads to non-Markovian random walks, which in turn 
generate correlations between halo formation history and large-scale environment \citep[e.g.,][]{ma10}.
\citet{desjacques08} shows that under the spherical collapse model, 
the correlations arising purely from this non-Markovianity are expected to 
be stronger for more massive halos, in disagreement with the trend seen in simulations.
Taking into account ellipsoidal collapse and the effects of large-scale environment on halo formation statistically
\citep{sandvik07} and dynamically \citep{desjacques08}, 
the age dependent assembly bias can naturally arise and be more apparent
for less massive halos. It is also noted by \citet{desjacques08} that
environmental density plays the determining role in the virialization redshift of halos,
and that the morphology of large-scale
structure contributes mostly to its scatter. 
However, given a fixed halo mass, the analytical model of \citet{desjacques08} 
predicts that the median formation redshift of halos decreases with increasing
environmental density, defined as the linear density fluctuation smoothed over $10h^{-1}\hbox{ Mpc}$. 
This is inconsistent with the results from simulations.  
\citet{hahn09} emphasize that for galactic-scale halos of fixed mass, 
an early formation epoch is closely related to a reduced mass-growth rate at late times that is mainly
due to the tidal effects of neighboring massive halos. As galactic halos are enhanced in filaments
near massive halos, the dependence of their formation epoch on environmental density is expected
\citep[see also, e.g.,][]{diemand07, wang07}. For cluster-scaled halos, however, \citet{dalal08} show that 
tidal effects from large-scale structure cannot play an important role in the environment
dependence of formation history. 

Gas physics allows more complicated interactions between galaxies and their environment. 
Galaxies are more subject to ram pressure \citep[e.g.,][]{frenk99} and shock heating \citep[e.g.,][]{mori00} 
in denser regions. Both the cooling efficiency and the gas compression rate are directly 
related to environmental density. It has been observed in hydrodynamical simulations that 
most cold flows are highly anisotropic and follow dark matter filaments \citep[e.g.,][]{keres05,dekel06}.

In this paper, we use SDSS DR7 data to examine the dependence of galaxy properties on their large-scale environment. 
We focus on the influence of the morphology of environment over and above the well-known dependence on environment density. 
This analysis may provide important constraints on the theories and hypotheses discussed above. 
It can also help clarify if large-scale environment should be included in the halo occupation distribution model
\citep[e.g.,][]{blanton07,tinker08}.  
The rest of the paper is organized as follows. In Section 2, we describe our measurements 
of large-scale tidal field with SDSS DR7.
In Section 3, we present our results.
Comparisons with results from simulations with semi-analytical modeling of galaxy formation 
are shown in Section 4. Section 5 contains discussions and our conclusions.
\section{Large-scale tidal field in SDSS}

\subsection{SDSS and NYU-VAGC}
The base data set adopted in this paper is SDSS DR7 \citep{york00,abazajian09}.
SDSS used a dedicated wide-field 2.5 m telescope \citep{gunn06} located at Apache Point Observatory to 
obtain images in the $u,g,r,i,z$ bands over an area of $\sim 10,000 \hbox{ deg}^2$ 
and spectra of $\sim 1.6$ million selected objects among which galaxies amount to $\sim 1$ million. 
The photometric data was calibrated through ubercalibration \citep{padmanabhan08}, 
which used the overlap between adjacent imaging runs to tie together the photometry of all imaging observations. 
Fibers arranged on plates $1^{\circ}.49$ in radius were assigned with an efficient tiling algorithm to 
spectroscopic targets selected with certain photometric criteria.  
Due to the fiber collision effect, only one spectroscopic measurement can be done 
for a group of targets with separations less than $55$ arsec.
Overlaps of tiles can help to reduce the number of missed targets. The overall incompleteness
for galaxies is $\sim 6\%$. Our test shows that this incompleteness together with $<2 \%$ incompleteness 
caused by mismeasurement of redshift in spectra and bright star blocking is negligible for our 
measurements of the large-scale tidal field.

Our analysis is based on the large-scale structure sample VAGC dr72 of New York University Value Added Catalogue 
(NYU-VAGC;\citet{blanton05a}). It is constructed from SDSS, FIRST, 2MASS, 2dFGRS and PSCz. The SDSS part
is updated for DR7. There are $\sim 2.5$ million galaxies in VAGC dr72. From VAGC dr72, we build up three volume-limited samples 
in which galaxies have absolute r-band magnitude brighter than $-18+5\log_{10}(h)$, $-19+5\log_{10}(h)$ 
and $-20+5\log_{10}(h)$ respectively, ($M_r$18, $M_r$19 and $M_r$20 hereafter).  We consider all 
spectroscopic galaxies with listed apparent magnitude 
brighter than r = 18.
All magnitudes mentioned in this paper are Petrosian magnitudes, and are K-corrected to redshift $z=0$.  
For $M_r$18, $M_r$19 and $M_r$20, the respective number of galaxies is $26760$, $53992$ and $106033$.
We mainly focus on the $M_r$20 sample because its volume is large enough for accurate measurements of 
the large-scale structure geometry and it does not suffer galaxy bias severely since $M_{*}$ in the luminosity 
function is a little brighter than $-20$. 
 
The VAGC provides us information on the size and the stellar mass for each galaxy, 
so that we can directly draw four parameters, concentration $C=R_{90}/R_{50}$, $(g-r)$ color, 
stellar mass $M_{*}$, and the surface stellar mass density $\mu_{*}=M_{*}/(2\pi R_{50}^2)$, 
where $R_{50}$ and $R_{90}$ are defined as the radii containing $50\%$ and $90\%$ of the Petrosian flux, respectively. 
In VAGC, calculations for the stellar mass $M_*$ are based on five-band photometric data and redshift. 
\citet{li09} compare this stellar mass measurement with spectroscopically based stellar mass measurements
by \citet{kauffmann03}, and show that the two results agree very well. 
We also add the amplitude of the 4000 $\AA$ break D$_{n}$(4000) into our analysis as a clear indicator of stellar population 
age, from the MPA/JHU SDSS data (http://www.mpa-garching.mpg.de/SDSS/). 
There are $105034$ galaxies in $M_r$20 that have counterparts in MPA/JHU data set.

\begin{figure}
\centering
\includegraphics[width=42mm]{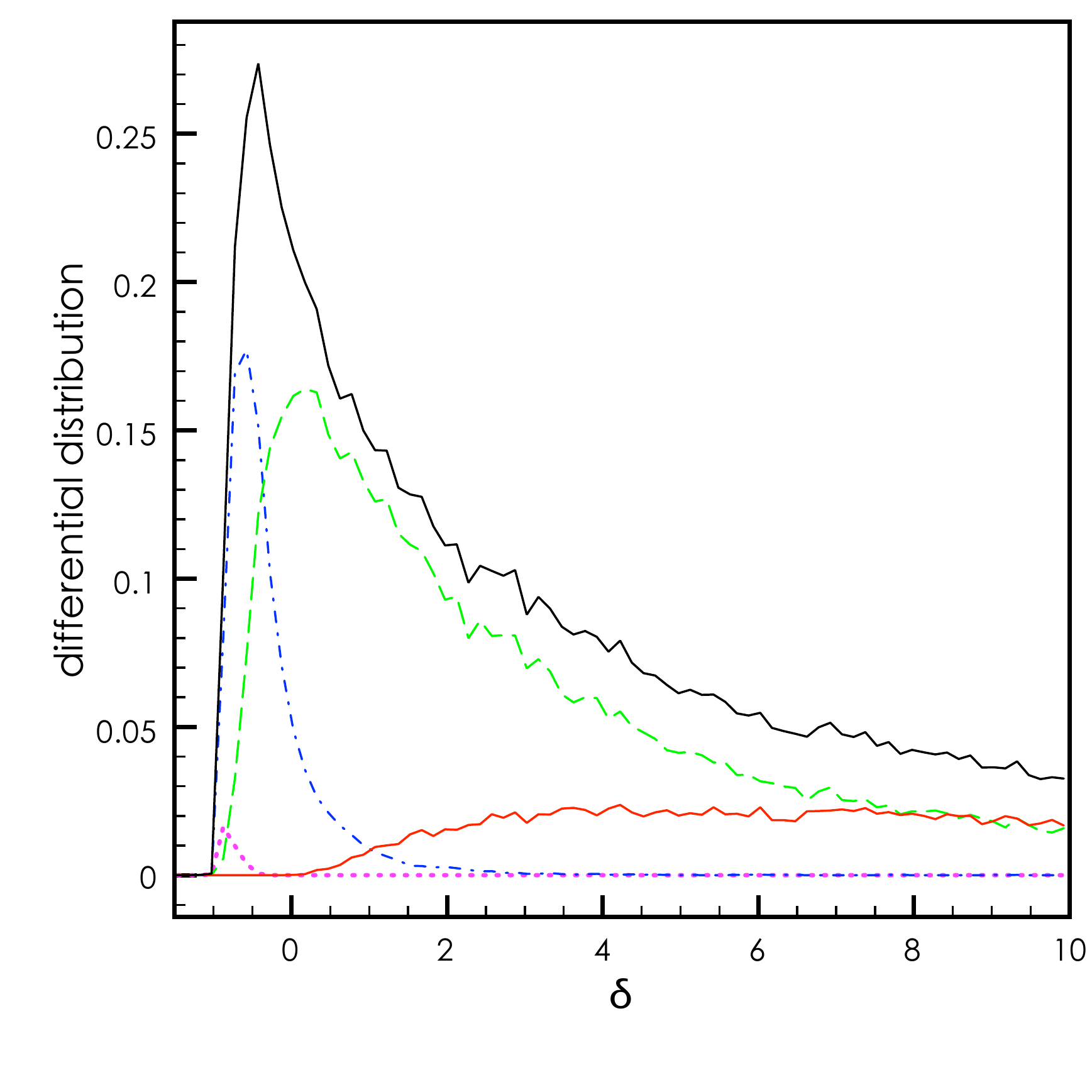}\includegraphics[width=42mm]{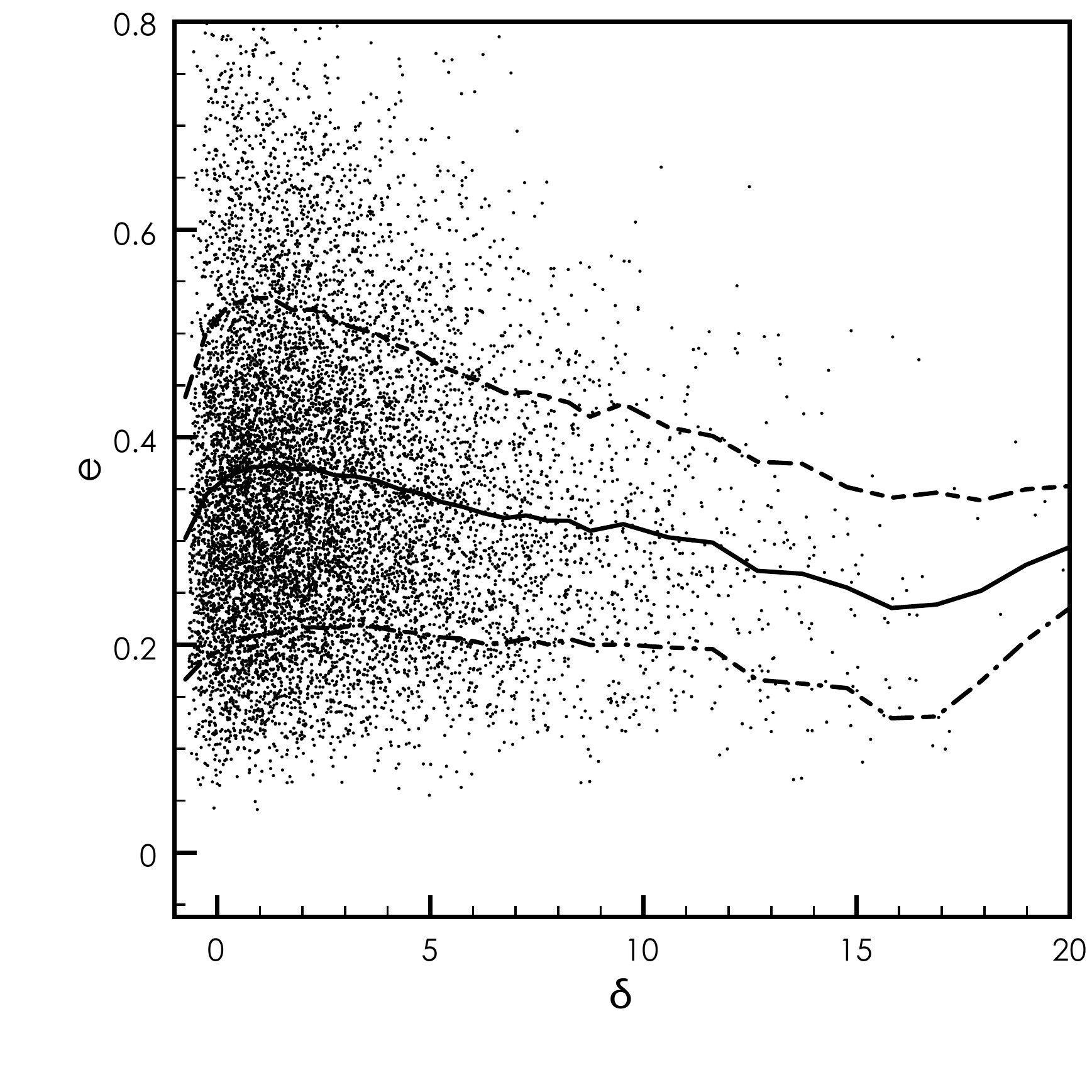}
\caption{Left panel: The environmental density distributions for galaxies in eigenvalue-sign-judged halos 
(red solid line), filaments (green dashed line), sheets (blue dot-dashed line) and voids (magenta dotted line). 
The upper solid line shows the environmental density distribution for the whole sample. The smoothing scale
for the environment of a galaxy is chosen to be the distance to its $3rd$ nearest neighbor. 
Right panel: Scatter plot of the environmental density $\delta$ and ellipticity $e$.  
The solid line is the average $e$ at different $\delta$ and the dashed lines show $\pm 1\sigma$ $e$ range
around the average $e$. }
\end{figure}

\begin{figure}
\includegraphics[totalheight=84mm,width=84mm]{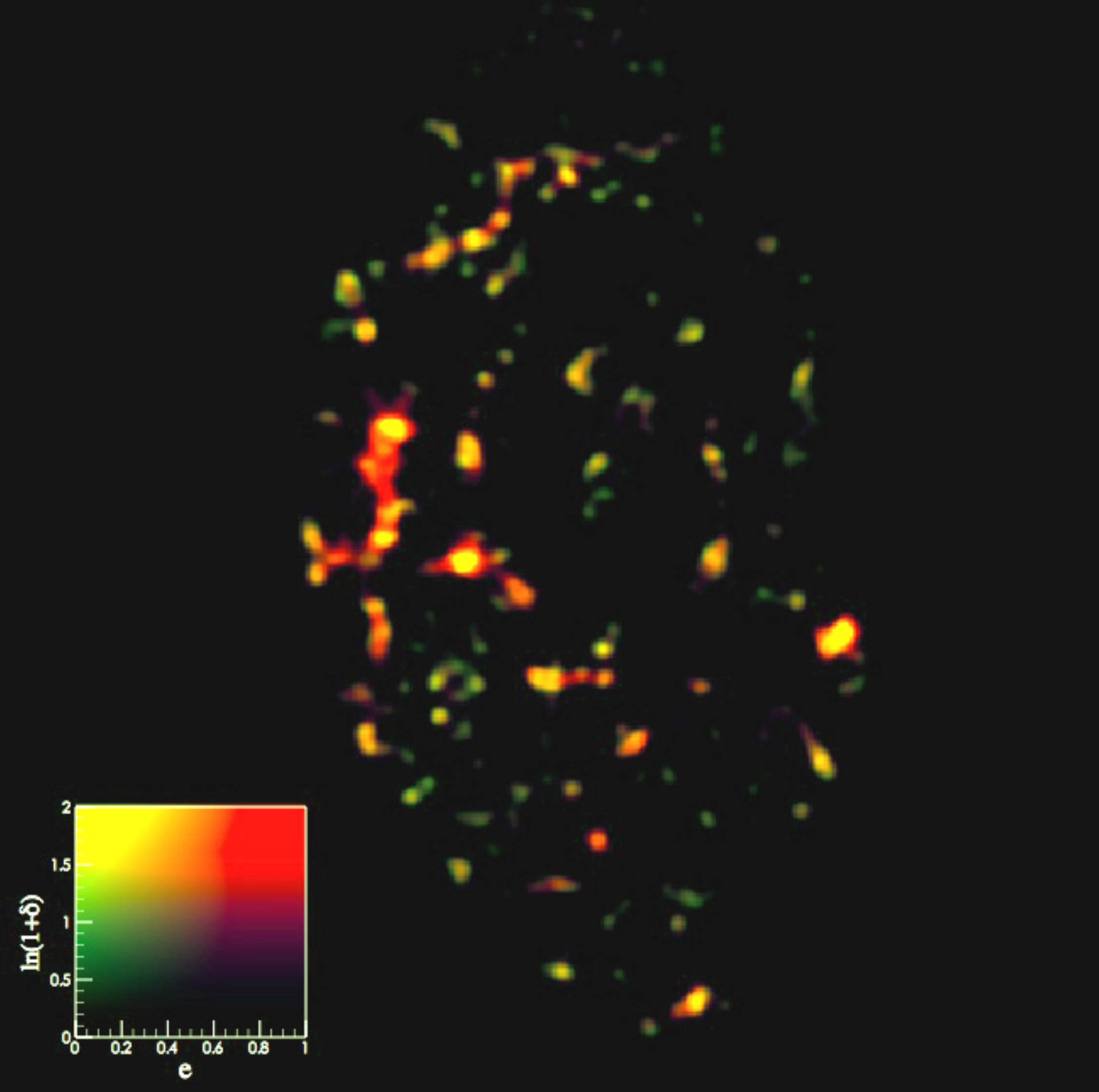}
\caption{Contours of environmental $[e,\ln(1+\delta)]$ for a slice of SDSS DR7. 
The environmental density and ellipticity are encoded with two-dimensional color scheme shown in the insertion. 
The smoothing scale is $3h^{-1}\hbox{ Mpc}$.}
\end{figure}

\subsection{Tidal environment measurements}

From a theoretical point of view, the tidal environment of a galaxy can be characterized by the
 second derivatives of the gravitational potential, 
$\partial _{i}\partial_{j} \phi$. The potential cannot, however, be measured directly,
and needs to be calculated through the distribution of galaxies in redshift space.
Such a measurement can be affected by, e.g., galaxy bias, redshift-space distortion and survey geometry. 
Here we describe our procedures to calculate the potential field, leaving detailed discussion to Section 4. 

For calculations of the potential field, we put the volume-limited sample into a large cubic box with 
side $500h^{-1}\hbox{ Mpc}$. Galaxies are assigned onto a $512^{3}$ uniform grid with CIC interpolation 
to give the number density of galaxies in each cell.
We remove galaxies from those cells that happen to cross the survey boundary,
and then assign the average number density of galaxies to all the empty cells. 
 Filling the regions outside the survey volume with the average number density 
allows large voids with size comparable to the survey volume to be identified properly.
This galaxy number density field is then smoothed with a Gaussian filter
with a smoothing scale $R_s$, and the potential field and further the tidal tensor field $\partial _{i}\partial_{j} \phi$
at each grid point are calculated from the smoothed density fluctuation field. 
The tidal environment of each galaxy is obtained by linearly interpolating this discrete tidal tensor field from the mesh to
its position. 

\begin{figure*}
\includegraphics[width=180mm]{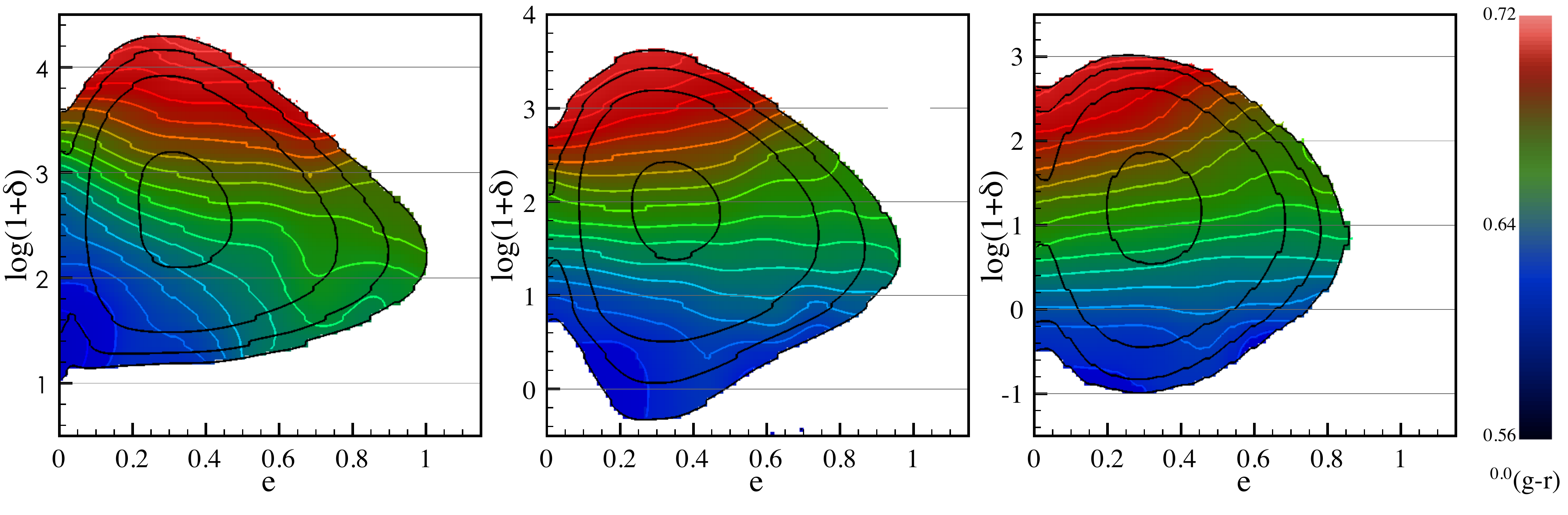}
\caption{Galaxy color contours in the ellipticity-density plane for the SDSS DR7. The fixed smoothing scales are $1h^{-1}\hbox{ Mpc}$ (left panel), 
$2h^{-1}\hbox{ Mpc}$ (middle panel) and $3h^{-1}\hbox{ Mpc}$ (right panel), respectively. The interval 
of the color contours is taken to be $0.05\sigma$. The black rings 
are the uncertainty contours of galaxy color with values 0.05$\sigma$, 0.1$\sigma$, 0.15$\sigma$ and 0.2$\sigma$ from inside out,
where $\sigma=0.135$ is the standard deviation in galaxy color for the whole $M_r$20 sample.
The horizontal black lines are plotted so that the behavior of correlation/anticorrelation with ellipticity can be seen.}
 \end{figure*}

\begin{figure*}
\includegraphics[width=180mm]{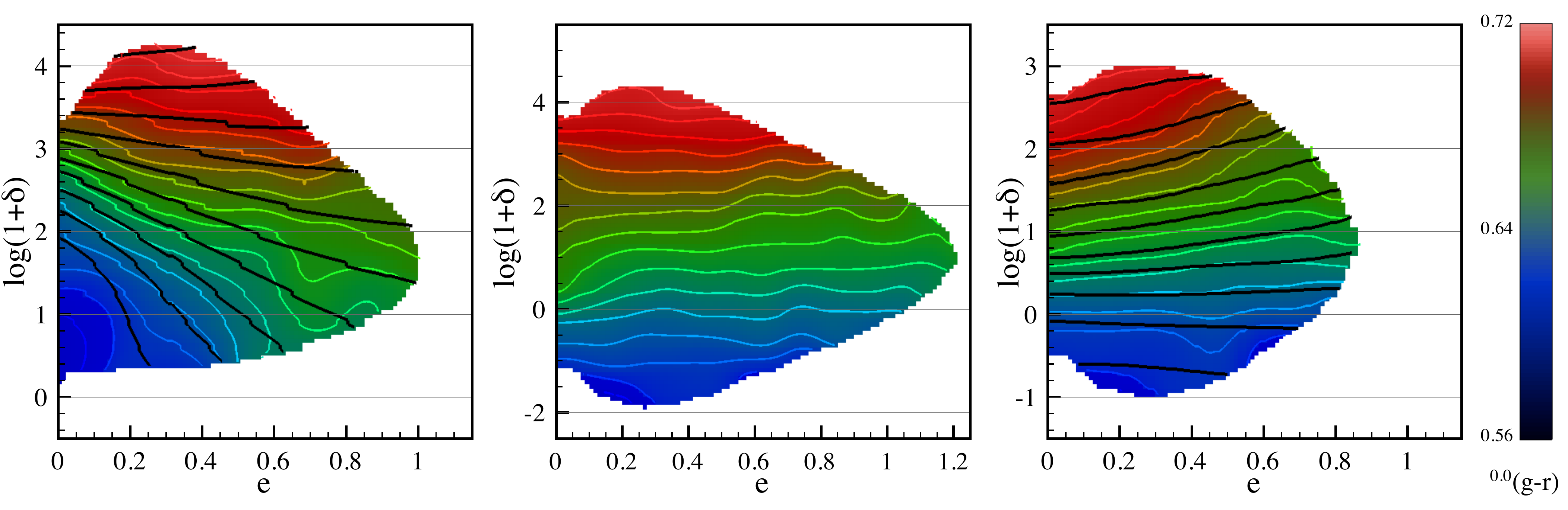}
\caption{Galaxy color contours similar to Figure 3. The smoothing scales are $1h^{-1}\hbox{ Mpc}$ (left panel),   
the `adaptive smoothing scale' (middle panel) and $3h^{-1}\hbox{ Mpc}$ (right panel), respectively. Here the 
`adaptive smoothing scale'  scale applied to smooth the environment of a galaxy is defined as the distance to its $3rd$ nearest neighbor.
The black contours shown in the left and right panels are the contours of $\ln(1+\delta_o)$ with
$\delta_o$ the environmental density measured by this same optimal adaptive smoothing.}
\end{figure*}

Without considering the specific orientation of the tidal tensor, its properties are fully characterized 
by its three eigenvalues $\lambda_1\ge \lambda_2\ge \lambda_3$. One way to classify the tidal environment of a galaxy
uses the signs of $\lambda_i \hbox{ }(i=1, 2, 3)$ \citep[e.g.,][]{hahn07, zhang09}. 
Specifically, one can define a point as (i) ``cluster'', if all three eigenvalues are positive; (ii)``filament'', 
if one eigenvalue is negative; (ii)``sheet'', if two eigenvalues are negative; and (iv)``void'', if 
all the three eigenvalues are negative. These definitions can show clearly the 
filament-node nature of large-scale structure at least in a two-dimensional cut. 
However, this classification is highly degenerate with a classification by density. 
As shown in the left panel of Figure 1, the overdensity distributions of points in these different categories are very different,
and thus the dependence of galaxy properties on such ``tidal environments'' may 
mainly reflect a density dependence. 
To address the dependence of galaxy properties on tidal properties
other than overdensity, we here describe the tidal environment of galaxies with parameters describing
ellipticity $e$ and prolateness $p$. 
Specifically, we define $e$ and $p$ as follows
 \[
   e=\frac{\lambda_{1}-\lambda_{3}}{3+\delta},      \qquad    p=\frac{\lambda_{1}+\lambda_{3}-2\lambda_{2}}{3+\delta},
  \]
where $\delta=\lambda_1+\lambda_2+\lambda_3$. We find that these definitions,
with the number $3$ added to the denominators, can minimize the correlations between $e$ ($p$) and $\delta$, as seen in 
the right panel of Figure 1. Figure 2 shows a 2-dimensional color rendering of $[e, \ln (1+\delta)]$ in a 2-dimension cut through the
$M_r$20 sample.
One can see that the cluster-filament-cluster structure is very well represented.
High density and high ellipticity bridges link high density and low ellipticity clusters. 
In the following analysis, we mainly focus on the $e$-dependence of galaxy properties, since
the $\delta$-dependence has been well explored in previous work.

\section{The Tidal dependence of galaxy properties}
\subsection{Tidal dependence of galaxy color}

As can be seen from Figure 2, giant filaments link clusters together. These filaments have similar 
densities to the outskirts of clusters. By introducing the additional parameter $e$, however, these two regions can be 
separated clearly. 

The main purpose of our study is to investigate if galaxy properties depend on the morphology of their 
large-scale environment in addition to the known density dependence. To do that, we consider
galaxy properties in the $[e,\ln (1+\delta)]$ plane, and analyze the correlations of these properties with $e$ and 
$\ln (1+\delta)$. Specifically, we divide the $[e,\ln (1+\delta)]$ plane into $128^2$ cells, 
where $e\in [0,1.2]$ and the typical range of $\ln (1+\delta)$ is $(-1,4)$ with its precise
range varying somewhat from one smoothing scale to another. 
We then extend the plane to a periodic $256^2$ grid by reflecting
the original cells with respect to $e=0$ and a line of the lowest value of $\ln (1+\delta)$.
At each cell, we calculate the total number of galaxies $N[e,\ln (1+\delta)]$ and the sum of the 
considered galaxy property $C[e,\ln (1+\delta)]$. Smoothing is then applied to both $C$ and $N$ with a Gaussian
smoothing function $G_{5}$, where subscript $5$ denotes that the smoothing scale is $5$ cells,
and the average $\langle C\rangle_s$ is obtained at each grid point by $\langle C\rangle_s=C_s/N_s$, where the subscript $s$
represents the smoothed quantities.  

To quantify the statistical significance of the $e$-dependence of galaxy properties, 
we need to estimate the statistical errors for the smoothed quantities properly.
Our sample contains $\sim 0.1$ million galaxies. When they are distributed onto $128^2$ cells in the $[e,\ln (1+\delta)]$
plane, we expect that Poisson fluctuations due to the limited number of galaxies associated with each cell
will dominate the statistical errors. We thus estimate the error in each cell as ${\sigma\sqrt{N_2}}/{N_1}$, 
where $\sigma$ is the standard deviation of the considered galaxy property over the whole sample, 
and $N_{1}$ and $N_{2}$ are the smoothed cell count with the smoothing functions $G_{5}$ and 
$G^{2}_{5}$, respectively. In Table 1, we list the values of $\sigma$ for different galaxy properties
for $M_r$20.

\begin{table}
 \caption{Average value and standard deviation of different galaxy properties for $M_r$20}
 \label{symbols}
 \begin{tabular}{@{}lcccccc}

  \hline
   Property  &  (g-r) & -$M_r$ & $D_{n}4000$  & $\mu_{*}$ & R90/R50        & $M_{*}$  \\
   \hline
   mean                  & 0.663 & 20.63 & 1.676 & 9.076 & 2.706 & 10.30\\
   $\sigma$   & 0.135 & 0.481 & 0.259 & 0.376 & 0.433 & 0.256\\
 \hline
\end{tabular}
\end{table}
 
Figure 3 shows the results for the color $(g-r)$, where $\sigma=0.135$ as derived from the sample $M_r$20 
and shown in Table 1. The three different panels correspond to three different 
smoothing scales to represent the large-scale environment, $R_s=1h^{-1}\hbox{ Mpc}$, $2h^{-1}\hbox{ Mpc}$
and $3h^{-1}\hbox{ Mpc}$ from left to right.
The $(g-r)$ values are shown in color in each panel with the scale shown to the right of Figure 3.   
For the superposed white contour lines, the interval between the adjacent lines is $0.05\sigma$. 
The inside-out black lines show the error contours of $0.05\sigma$, $0.1\sigma$, $0.2\sigma$ and 
$0.3\sigma$, respectively.

The dependence of $(g-r)$ on the density $\delta$ is clearly seen in Figure 3.
The slope of the $(g-r)$ contours reflects whether the ellipticity $e$ also affects the galaxy color
in addition to $\delta$. We can see a weak correlation between $(g-r)$ and $e$, 
and this correlation is scale-dependent. For $R_s=1h^{-1}\hbox{ Mpc}$, an anti-correlation is seen. 
This reverses with increasing $R_s$, and is positive for $R_s=3h^{-1}\hbox{ Mpc}$ and larger. 
For $R_s=2h^{-1}\hbox{ Mpc}$, the correlation is nearly null.  

\subsection{Scale-dependence of environmental effects} 

From Figure 3, we see that galaxy colors depend strongly on environmental density $\delta$
and weakly on environmental ellipticity $e$. This behavior are scale-dependent.
The question arises then what smoothing scale one should adopt in order to reveal the 
most fundamental dependence of galaxy properties on their environment.  

Previous studies have shown that galaxy properties are influenced by their environmental densities
mainly locally \citep{kauffmann04,blanton07,park07}. On the other hand, if we simply smooth the environments 
on a scale larger than expected for any physical impact,
 we may still see an apparent environment-density dependence
which arises merely due to the correlation between 
the environmental density on the considered smoothing scale and that on the scale where
environmental effects are the strongest. This can also apply to the apparent $e$-dependence
of galaxy properties shown in Figure 3. 

The scale-dependent switch from anti-correlation to positive-correlation with $e$ for galaxy colors seen in 
Figure 3 indicates to us that the best smoothing scale should be around the transition scale
where no $e$-correlations are observed. Furthermore, this transition scale should correspond
to the scale where the density dependence of galaxy properties is the strongest.
In Appendix A, we discuss further this scale-dependence of the environmental-density influence on galaxy properties.
We find that the scale where the environmental-density dependence of galaxy properties is the strongest 
is somewhat different for high and low density regions with the values
of $\sim 1.5 h^{-1}\hbox{ Mpc}$ and $2.5 h^{-1}\hbox{ Mpc}$, respectively. This difference suggests 
that an adaptive smoothing scale may be more appropriate. A natural choice is
the characteristic distance between the considered galaxy and its neighbors. Our analysis show that
for $M_r$20 sample, the distance to a galaxy's $3rd$-nearest-neighbor is an optimal adaptive smoothing scale, 
which has a distribution peaked at $2 h^{-1}\hbox{ Mpc}$  and is 
close to the peak scale where galaxy properties and environment density have the strongest correlation in both
high and low density regions. 
For $M_r$19 and $M_r$18 samples, this optimal scale should correspond to the distance to a galaxy's $6th$-nearest-
neighbor and $8th$-nearest-neighbor respectively, due to the fact that the galaxy density is higher.

In Figure 4, we show $(g-r)$ contours as in Figure 3. The right and left panels are for fixed
smoothing scales of $1 h^{-1}\hbox{ Mpc}$ and $3 h^{-1}\hbox{ Mpc}$, respectively.
The middle panel shows results using the $3rd$-nearest-neighbor distance as an adaptive smoothing scale.
In comparison with the middle panel of Figure 3, the null-dependence on $e$ is more cleanly seen 
for the adaptive smoothing. To investigate if the dependence on $e$ seen in the left and right panels 
of Figure 4 (the same as those of Figure 3) is physical or due to the correlations between $(e,\delta)$ at
the considered scale and $\delta$ at the `optimal adaptive smoothing scale', we calculate the latter correlations. Specifically,
we define $(e,\delta)$ at the considered smoothing scale and at the optimal adaptive smoothing scale as $(e_a, \delta_a)$ 
and $(e_o, \delta_o)$, respectively. We compute the $\delta_o$-contours in the $(e_a, \delta_a)$ plane,
which are shown as black lines in the left and right panels of Figure 4. It is seen that the $\delta_o$-contours
align with the $(g-r)$-contours (white lines) very well. This demonstrates that the 
observed $e$-dependence is mainly attributable to the correlation between $(e_a, \delta_a)$ and $\delta_o$.
Some deviation between the $(g-r)$-contours and the $\delta_o$-contours are seen
in high density regions. These are, however, insignificant compared to the statistical errors. 
Together with the null-correlation with $e$ for the optimal adaptive smoothing, we conclude that
no significant physical dependence of galaxy colors on $e$ is detected by our analysis.  

\subsection{Other galaxy properties}

We have studied the dependence of galaxy color on $(e,\delta)$ carefully. Here we 
examine the tidal dependence of other galaxy properties.

Galaxy properties can be divided into three classes: (1) star-formation-related properties such as galaxy color 
and D$_{n}$(4000); (2) morphology-related properties such as concentration; and (3) extensive quantities such as the
total stellar mass $M_{*}$ and galaxy size. Previous studies have shown that star-formation-related properties 
depend directly on the environmental density \citep{kauffmann04,blanton05b,christlein05}, 
but morphology-related ones are not independently correlated with environmental density \citep{park07,ball08,bamford08}. 

Here we analyze the $(e,\delta)$-dependence of different galaxy properties. 
Figure 5 presents the environmental dependence of $(M_*, \hbox{D}_{n}\hbox{4000})$.
The left panel shows a scatter plot of $(M_*, \hbox{D}_{n}\hbox{4000})$. The 
red lines are contours of constrained mean value of $\ln(1+\delta_o)$. From the scatter plot, we see that 
$\hbox{D}_{n}\hbox{4000}$, which reflects stellar population ages and therefore star formation histories, correlates
with total stellar mass $M_*$. The directions of the red contours indicate that 
both depend on environmental density as measured by our optimal adaptive smoothing. In the right panel
of Figure 5, we show the $(e,\delta)$ dependence for $\hbox{D}_{n}\hbox{4000}$ given the
stellar mass $M_*$. Specifically, the contours of the deviation $\hbox{D}_{n}\hbox{4000}-\langle\hbox{D}_{n}\hbox{4000}\rangle$
are shown, where $\langle\hbox{D}_{n}\hbox{4000}\rangle$ is the average value of $\hbox{D}_{n}\hbox{4000}$ in 
stellar mass bins. The environmental-density dependence of the contours
are clearly seen, which, in agreement with the trend seen in the left panel, demonstrates the extra 
density dependence of $\hbox{D}_{n}\hbox{4000}$ in addition its dependence of $M_*$. 
On the other hand, no dependence on the shape parameter $e$ is detected. 

In Figure 6, we show a similar analysis for the concentration parameter $C=R_{90}/R_{50}$.
The scatter plot of $(\hbox{D}_{n}\hbox{4000}, C)$ is shown in the left panel with the $\ln(1+\delta_o)$ contours
superposed on it. Two clumps are seen in the scatter plot with one at low value of 
$\hbox{D}_{n}\hbox{4000}$ and small value of $C$, and the other at high $\hbox{D}_{n}\hbox{4000}$ and large $C$.
The lower-left clump corresponds to galaxies with recent star formation,
and the upper-right clump is associated with galaxies with quenched star formation.
The $\ln(1+\delta_o)$ contour lines are nearly vertical, showing no detectabe independent 
correlation between $C$ and the environmental density. In the right panel of Figure 6, 
we show the $C-\langle C\rangle$ contours in the $[e_o,\ln(1+\delta_o)]$ plane, where $\langle C\rangle$ is the average
value of $C$ given $\hbox{D}_{n}\hbox{4000}$. The chaotic pattern seen here reflects
the lack of any environmental dependence of the concentration $C$ other than its $\delta_o$ dependence
through the correlations with $\hbox{D}_{n}\hbox{4000}$. 

\begin{figure}
\includegraphics[width=42mm]{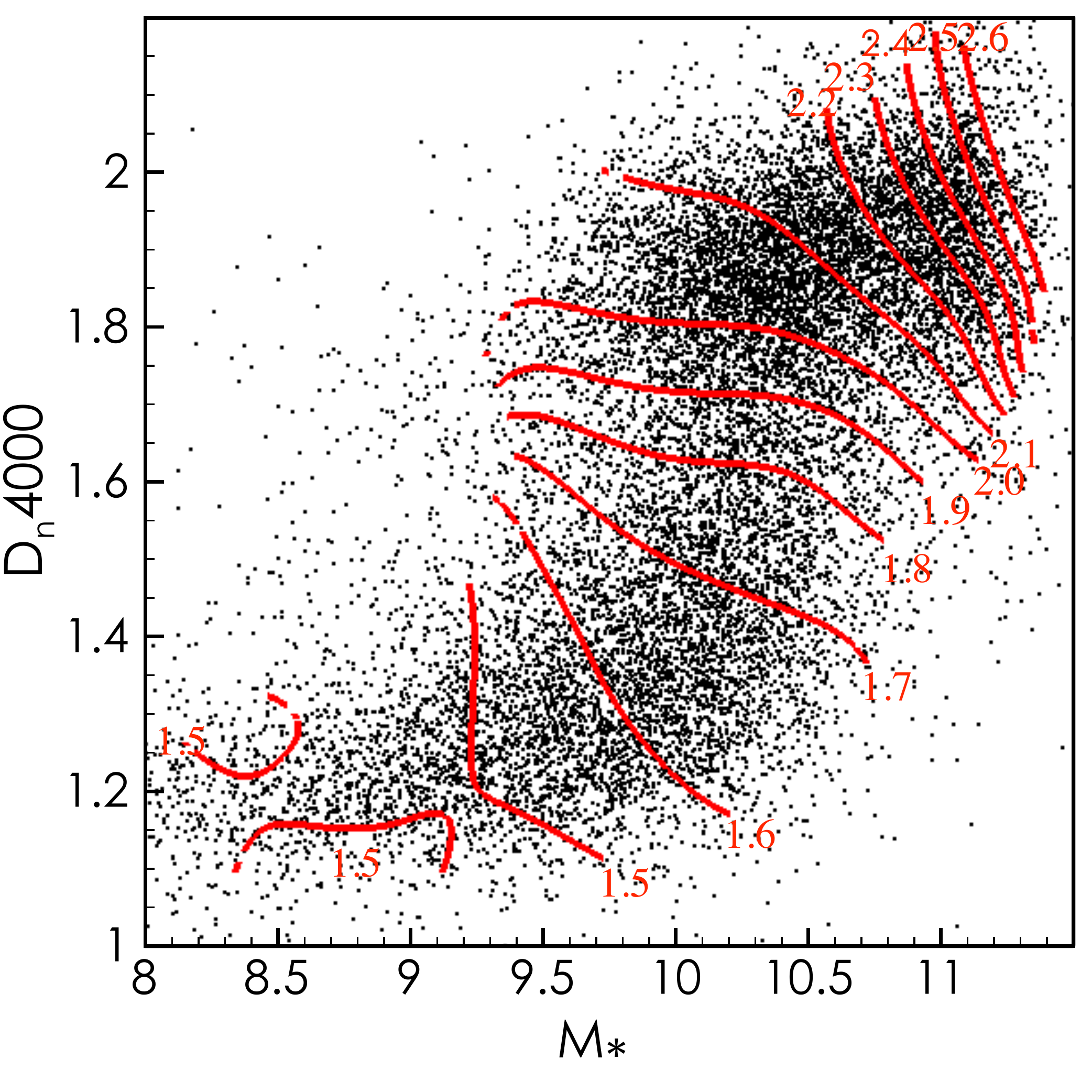}\includegraphics[width=42mm]{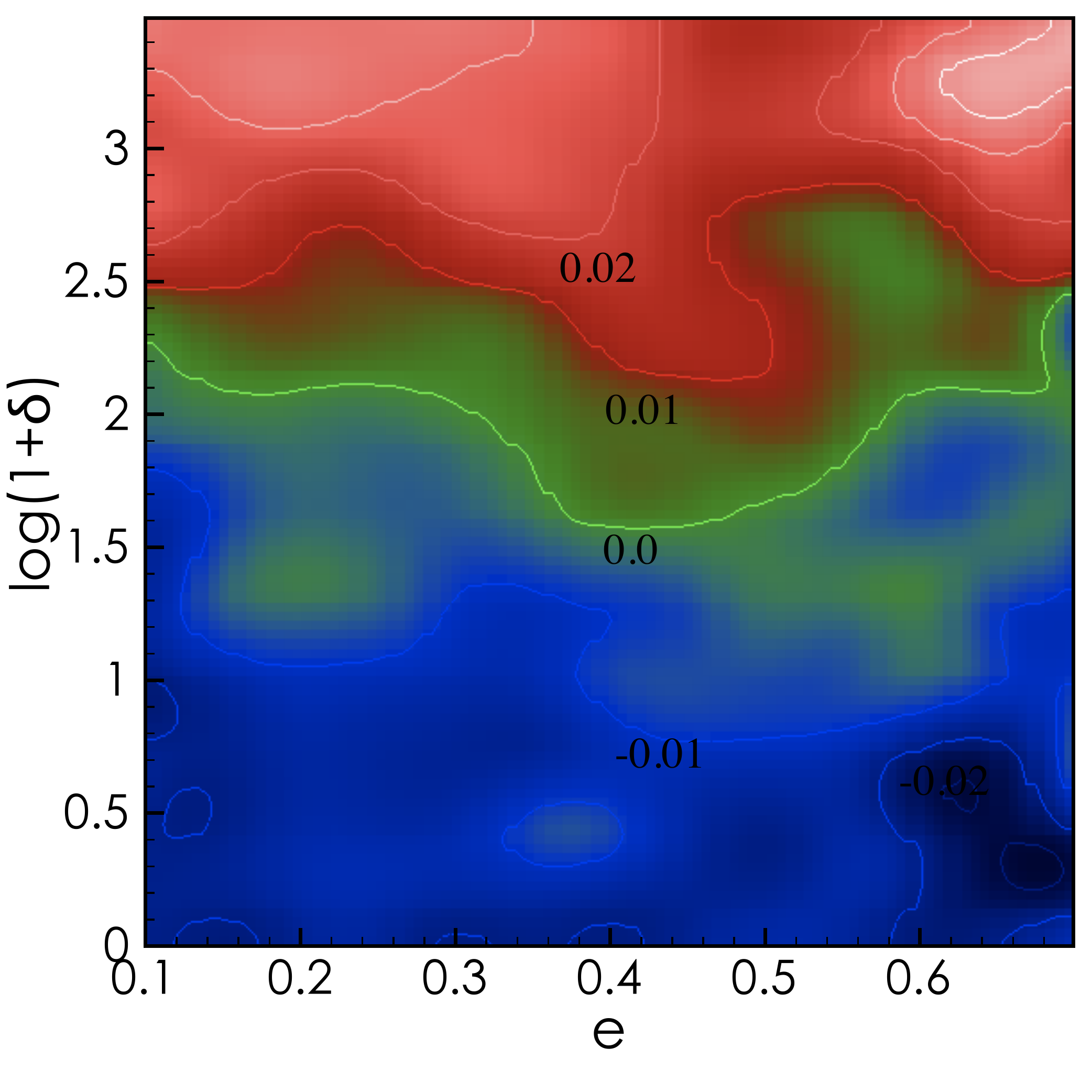}
\caption{Left panel: The scatter plot of ($M_*$, D$_{n}$(4000)). The $\ln(1+\delta_o)$ contours
are shown in red. Right panel: The contours of D$_{n}$(4000)-$\langle$D$_{n}$(4000)$\rangle$, 
where $\langle$D$_{n}$(4000)$\rangle$ is the average value of D$_{n}$(4000) given stellar mass.}
\end{figure}

\begin{figure}
\includegraphics[width=42mm]{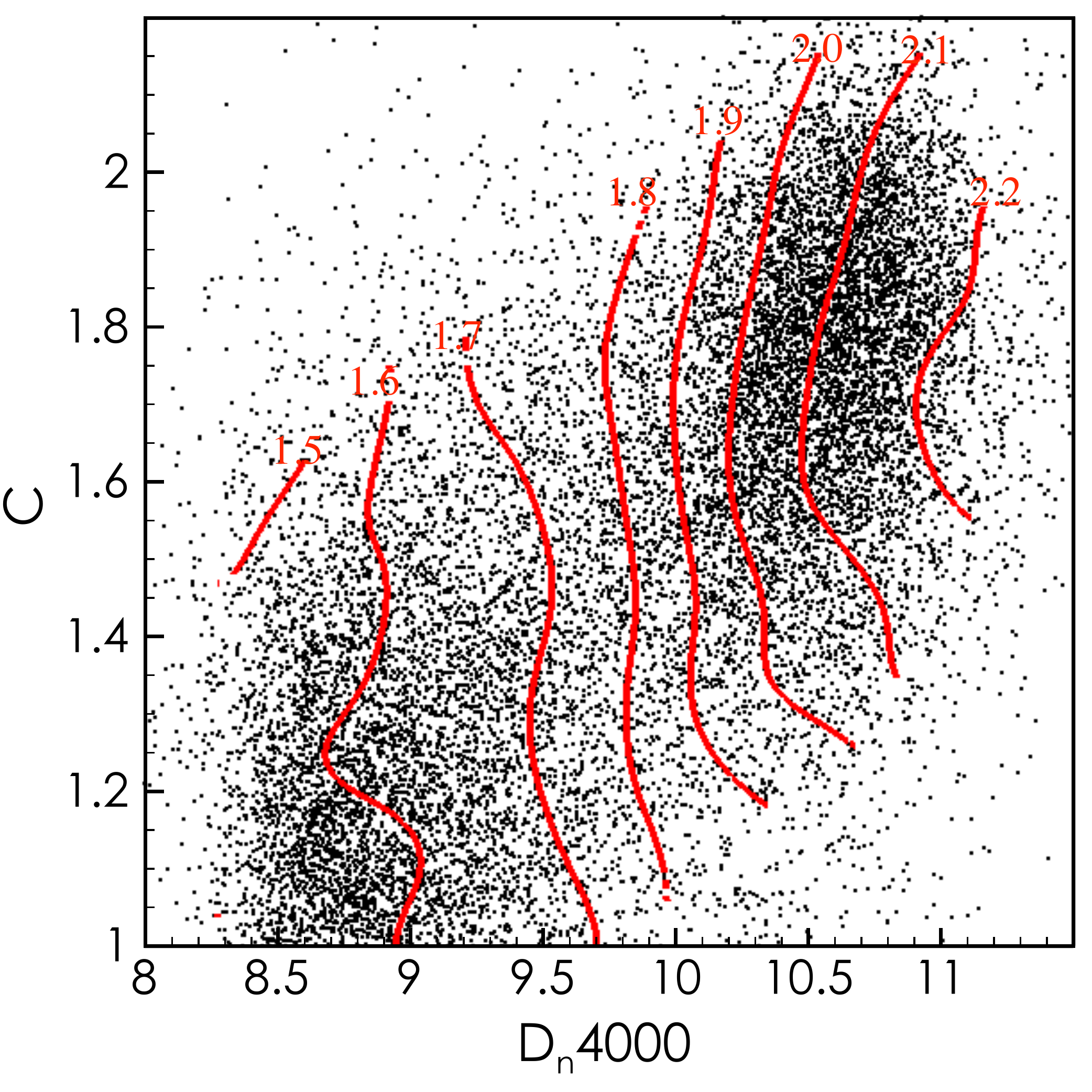}\includegraphics[width=42mm]{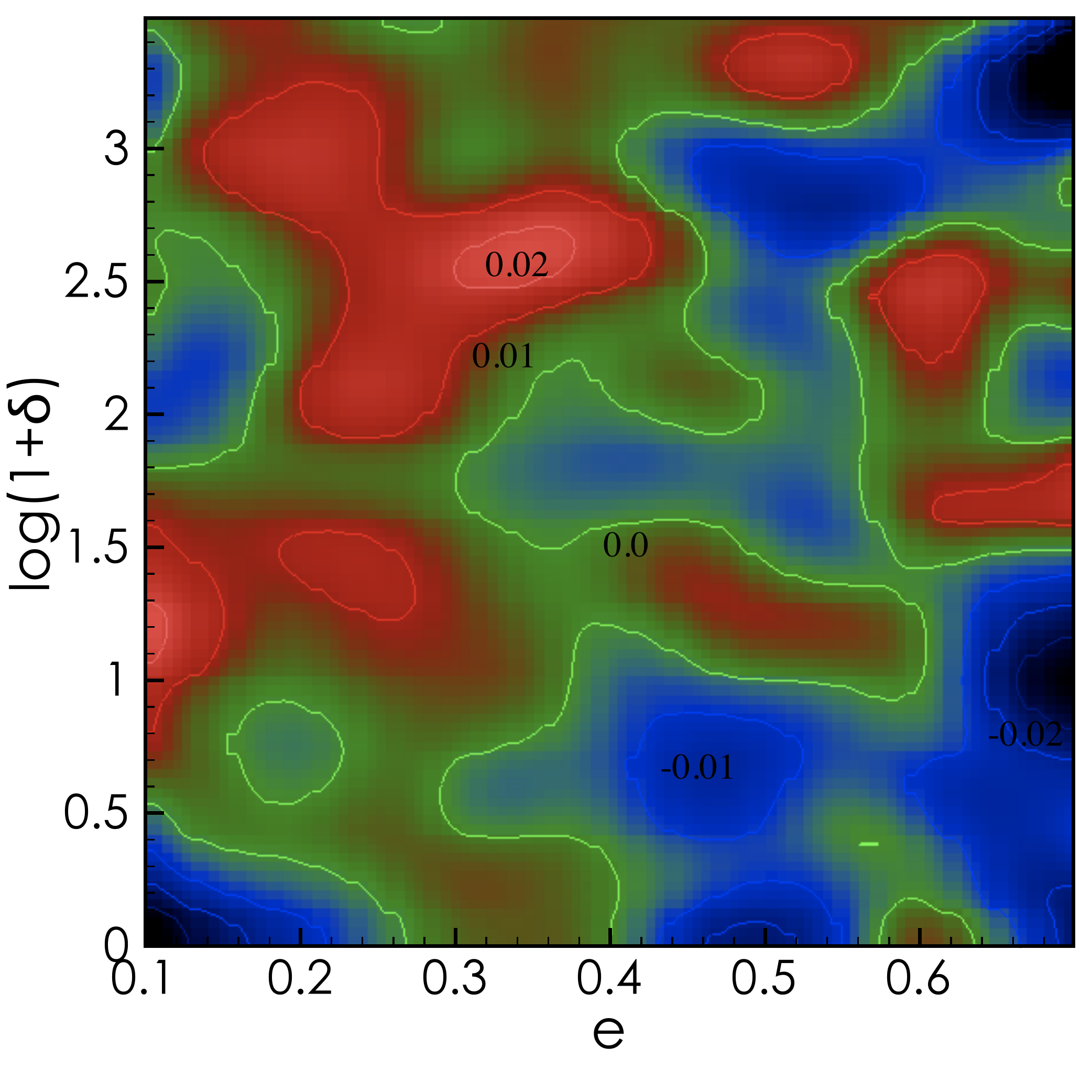}
\caption{Left panel: The scatter plot of (D$_{n}$(4000), $C$). The $\ln(1+\delta_o)$ contours
are shown in red. Right panel: Contours of $C-\langle C\rangle$, where $\langle C\rangle$ is the average value of $C$ given D$_{n}$(4000).}
\end{figure}

\begin{figure}
\includegraphics[width=84mm]{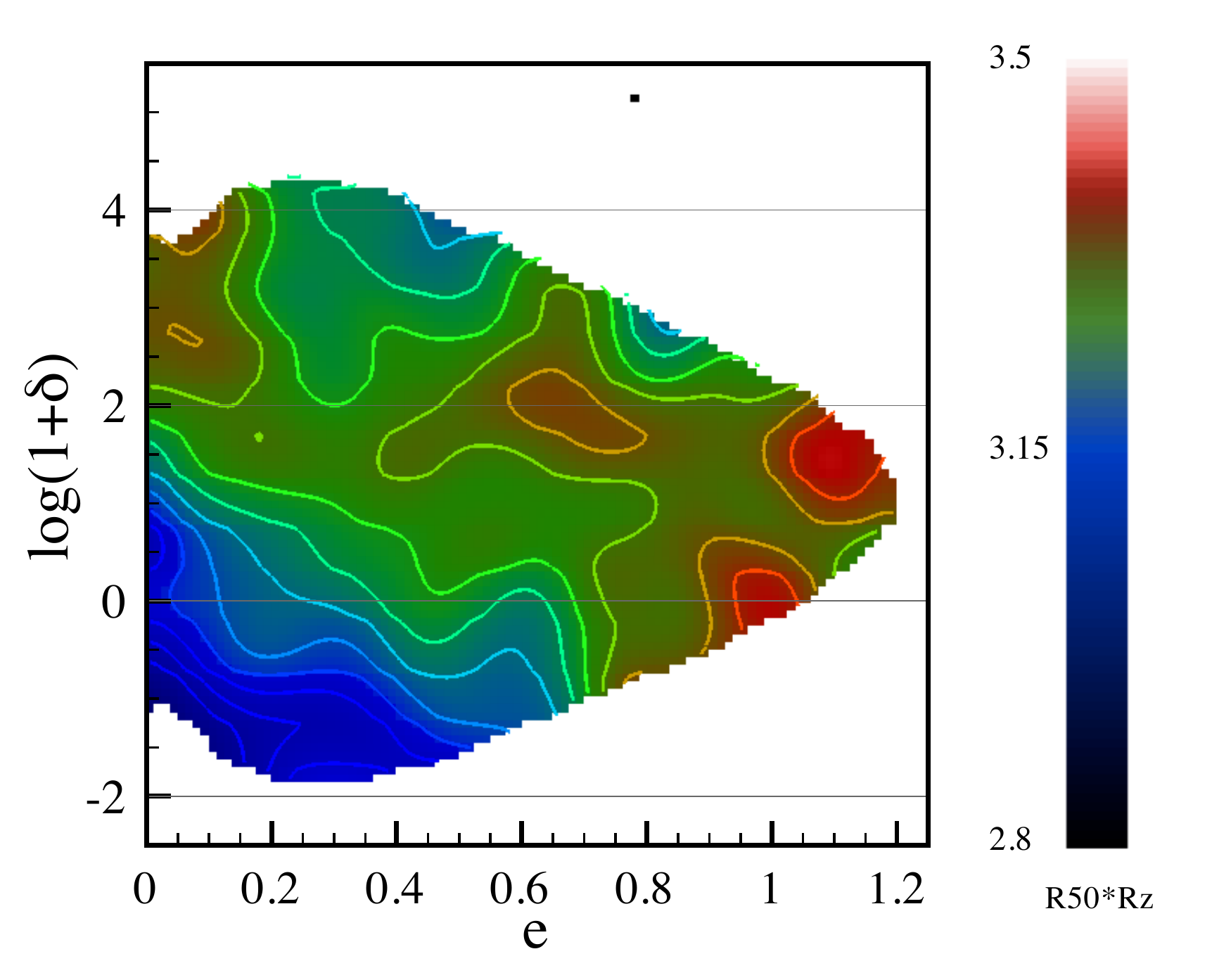}
 \caption{Contours of galaxy size defined as $R_{50}*R_z$ in unit of   for SDSS.
The boundary is the 0.2$\sigma$ uncertainty contour, where $\sigma$ is the standard deviation of 
galaxy size for the whole sample. The optimal adaptive smoothing is applied.}
\end{figure}

We also analyze the environmental dependence of galaxy size defined as $R_{50}$*d in unit of h$^{-1}$kpc, where
d denotes the distance to a galaxy. The results are shown in Figure 7, again using our optimal adaptive smoothing.
It is seen that in high density regions, galaxy size has no detectable correlation with either environmental density or
ellipticity. In low density regions, on the other hand, there is a notable correlation between galaxy size 
and $e_o$ in addition to its correlation with $\delta_o$. We will show in section 4.3 that in low density regions, 
the survey boundary effects can bias the $\delta_o$ and $e_o$ measurements
to lower values. This can induce an artificial correlation with $e_o$, which is considerable for small-volume samples
such as $M_r$18. For $M_r$20, however, the survey volume is large and the boundary effects are 
insignificant. Thus the correlations seen in the lower left region of Figure 7 may indicate
that both the density and the ellipticity of the environment, can affect the galaxy size.
Galaxies are systematically smaller in lower $(\delta_o, e_o$) regions. 
Due to the still limited statistics of SDSS, the environmental $e_o$ dependence of galaxy size
needs to be investigated further with future large surveys that can provide much improved statistics.

Our analysis so far has measured ellipticity and density on the same scale. The optimal adaptive smoothing 
is chosen to be the distance from a galaxy to its 3rd-nearest neighbor. This scale has a distribution peaked 
between 1.5h$^{-1}$Mpc and 2.5h$^{-1}$Mpc, depending on environmental density. In order to show the influence
of larger scales, especially the influence of the large filaments linking cluster as seen in Figure 2, we also analyze 
the environmental dependence of galaxies properties on ellipticity and density measured on two different scales. 
We fixs the smoothing for density to the optimal adaptive smoothing and we change the smoothing scale for ellipticity. For no smoothing 
scale larger than the 2h$^{-1}$Mpc do we 
see a correlation of galaxy properties with environment ellipticity. We thus conclude that large scale structure beyond galaxy groups
has no additional influence on individual galaxies. 

We have so far focused our analysis on the ellipticity $e$ of large-scale environment. 
However, we have also looked for the dependence of galaxy properties on the environmental prolateness $p$. 
We find that smoothing according to the distance to the $3rd$-nearest-neighbor of a galaxy, 
all of the considered galaxy properties are independent of $p$.

\section{Comparison with a semi-analytic model}
In semi-analytic models of galaxy formation (SAM), our best understanding on dark matter halo
formation is combined with the complex baryonic physics associated with galaxy formation as represented by
physically/observationally motivated prescriptions involving a set of parameters. 
While earlier SAMs were based on using the extended Press-Schechter theory to construct dark matter halo merger trees,
modern SAMs take full advantage of high resolution N-body simulations to follow the formation and evolution of
dark matter halos. Galaxies are then formed within these halos by modeling physical processes such as cooling, condensation,
star formation, supernova and AGN feedback.
Here we analyze the SAM galaxy catalog from \citet{lucia06}, for which the dark matter distribution was
taken from the Millennium Simulation \citep{springel05}. We compare the results with those from the SDSS. 
To test observational effects, we also construct mock SDSS DR7 catalogs from the SAM data \citep{li07} .
Specifically, we address the following issues 
\begin{enumerate}  
  \item a comparison of the environmental-density dependence of galaxy properties in the SAM and the SDSS;
  \item a comparison of the behavior of the tidal dependence of galaxy properties;
  \item the influence of various observational effects, such as redshift-space distortion and survey geometry
              on the results of our tidal analysis.
\end{enumerate}

\begin{figure}
\includegraphics[width=78mm]{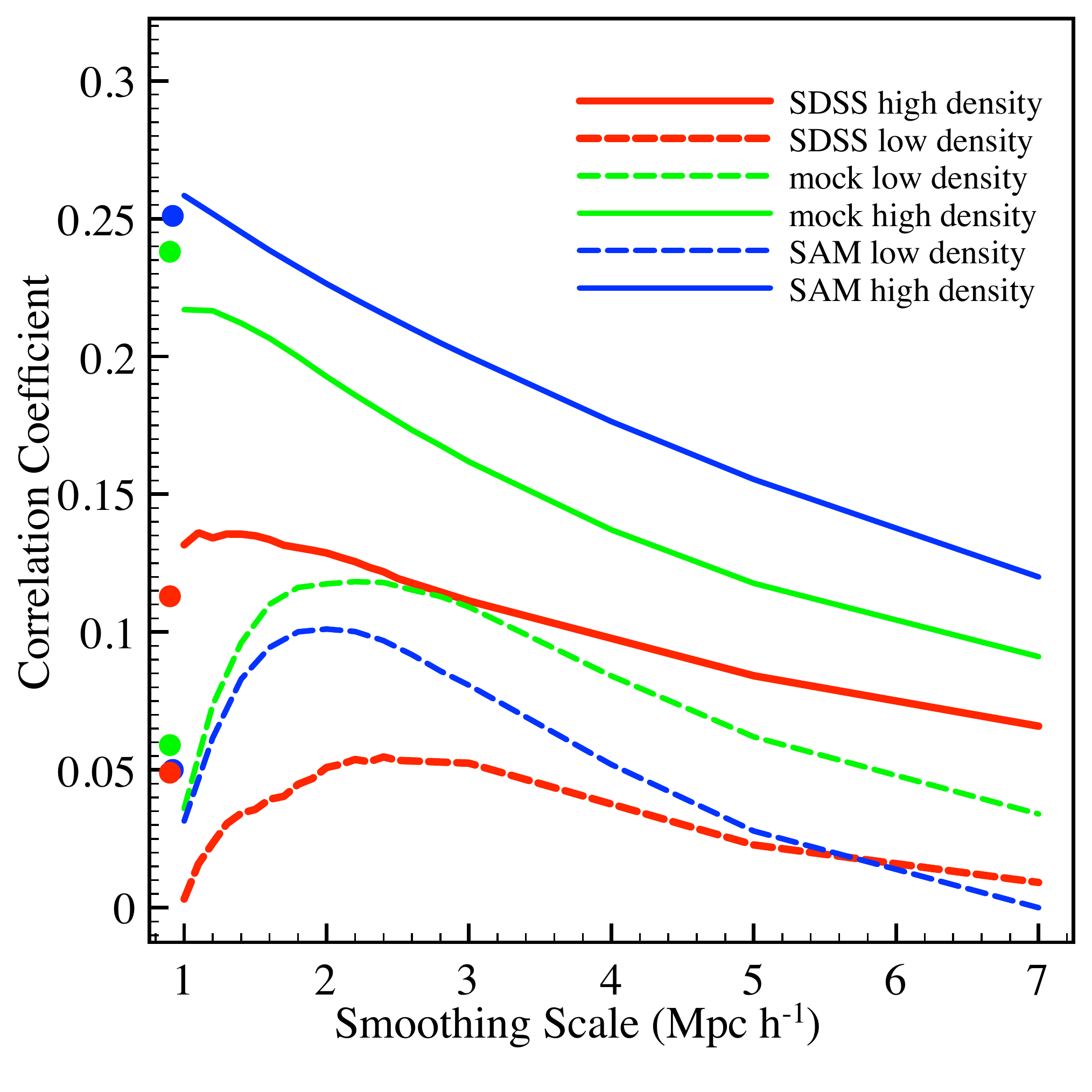}
\caption{The color-density correlation coefficient as a function of smoothing scale.
The red, blue and green lines give results for SDSS, SAM, and mock catalog, respectively. 
The solid and dashed lines are the results for galaxies in high and low density regions, respectively (see Appendix A). 
The filled circles of different color represent the results for optimal adaptive smoothing of the corresponding 
samples.
}
\end{figure}

\begin{figure*}
\includegraphics[width=180mm]{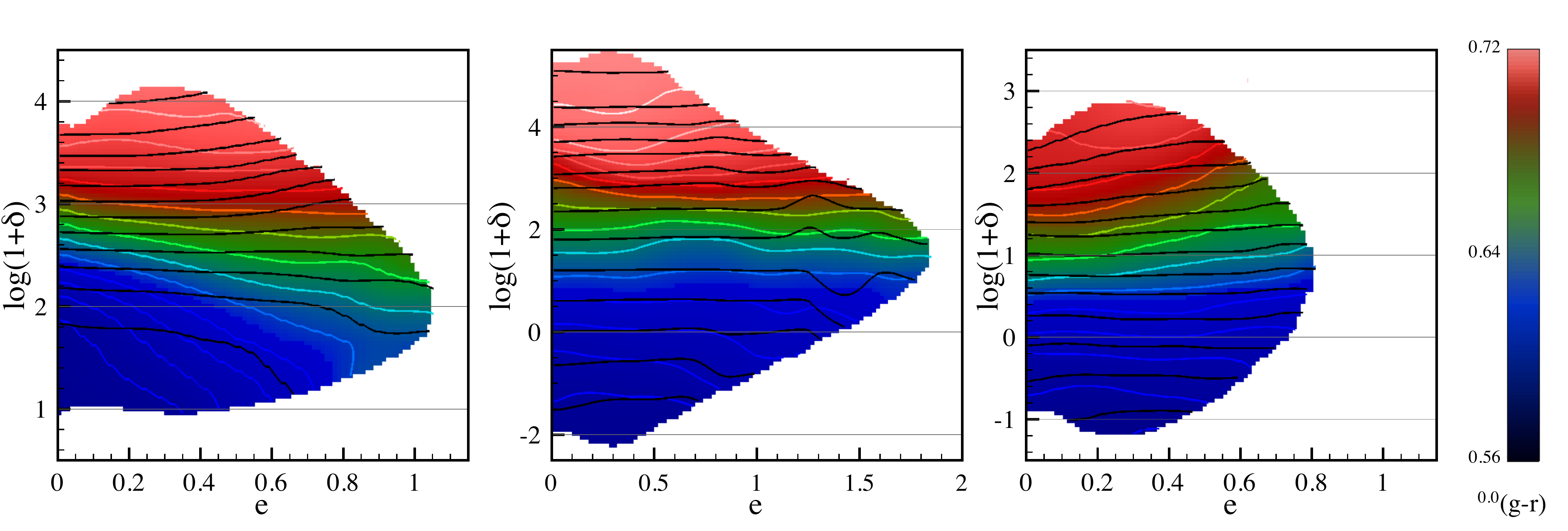}
\caption{Galaxy color contours similar to Figure 4, but for the SAM.
Here the black contours are the color contours for mock data. 
The smoothing scales are $1h^{-1}\hbox{ Mpc}$ (left panel), distance to the $3rd$-nearest-neighbor (middle panel) and 
$3h^{-1}\hbox{ Mpc}$ (right panel), respectively. The outer boundary is the 
uncertainty contour of galaxy color with the value of 0.1$\sigma$, where $\sigma=0.157$ is 
the standard deviation of the galaxy color distribution for the whole SAM sample.}
\end{figure*}

\subsection{Scale dependence of the color-density relation in the SAM}

To analyze the scale dependence of the color-density relation, we calculate the correlation coefficient 
between galaxy color $(g-r)$ and environmental density $\delta$ at different smoothing scales.
The correlation coefficient is defined as 
\[
r={\langle\Delta (g-r) \Delta \delta_o\rangle\over \sqrt{\langle[\Delta (g-r)]^2\rangle}\sqrt{\langle(\Delta \delta_o)^2\rangle}},
\]
where $\Delta (g-r)=(g-r)-\langle(g-r)\rangle$ and $\Delta \delta_o=\delta_o-\langle\delta_o\rangle$.
The results are shown in Figure 8. We divide galaxies into two groups of high (solid lines) and 
low densities (dashed lines) as described in Appendix A (the green lines in Figure A1). 
The red, blue and green lines are, respectively, for the SDSS, the full volume of the SAM,
and for the mock SDSS.

First, we see similar trends for the scale dependence of the color-density correlation 
in the SAM and the SDSS data. The SAM `optimal smoothing scale' for the strongest correlation is also consistent
with that of SDSS, which is $\sim 1h^{-1}\hbox{ Mpc}$ for high density regions and 
$\sim 2h^{-1}\hbox{ Mpc}$ for low density regions. On the other hand,
the amplitude of the correlation coefficient is significantly larger for the SAM than for the SDSS. 
The SAM predicts an environment-density dependence of galaxy colors  stronger than is observed,
in accordance with other studies \citep[e.g.,][]{springel05,coil08,font08,cowan08, guo11}. 
We note that while the tidal interactions of dark matter halos are naturally accounted for in the SAM,
tidal disruption of galaxies is not. For massive clusters, 
however, it is expected that their strong tidal forces may destroy galaxies \citep[e.g.,][]{henriques08}.
As those satellite galaxies are mostly red, such disruptions can decrease the level of clustering of red galaxies,
and therefore reduce the discrepancies in color-density relation between the SAM and the observations
in high density regions. Further, this can also alleviate the overly-fast-growth problem for central galaxies in the SAM
\citep{wang10}. It has also been argued that hot gas in satellite galaxies may not be 
immediately stripped away by ram pressure when they merge into a larger system as often assumed in SAMs. 
Therefore a reservoir of gas can be replenished to maintain the star formation for a 
relatively long period of time \citep[e.g.,][]{font08, coil08, wienmann06, guo11}.
In low density regions, the SAM also predicts stronger color-density correlations than seen in the SDSS, which reflects the smaller
color spread in the SAM compared to the SDSS. 

The green lines in Figure 8 are the results from the SDSS DR7 mock catalogue constructed from SAM data.
The differences between the green and blue lines reflect observatioal effects which will be discussed
in \S 4.3.

\subsection{Tidal dependence of galaxy properties in SAM}

To study the tidal dependence of galaxy properties, we perform the same analysis for the SAM data as shown in \S 3.2 for 
the SDSS.
The results are shown in Figure 9. The black lines here represent results from the mock catalogue, and will be discussed
in \S 4.3. The three panels correspond to the smoothing with $R_s=1h^{-1}\hbox{ Mpc}$,
with the $3rd$-nearest-neighbor distance, and with the $R_s=3 h^{-1}\hbox{ Mpc}$. It is seen that the contour behaviors 
are qualitatively similar to those of Figure 4 for the SDSS. In the SAM, the distance to the $3rd$-nearest neighbor
also corresponds to the 'peak scale' of the color-density dependence, at which, the tidal $e$-dependence of 
galaxy properties is nearly null. The density $\delta_o$ at the optimal adaptive smoothing scale plays the dominant role for the environmental
effects. The weak negative/positive $(g-r)$-$e$ correlations at smaller/larger smoothing scales
again reflect correlations between $(e,\delta)$ at the smoothing scale and $\delta_o$. 

A notable difference between Figure 9 and Figure 4 is the density dependence of color, which is significantly 
stronger in Figure 9 than that shown in Figure 4. This is consistent with the result shown in Figure 8, which also show
that the SAM predicts a considerable stronger color-density correlation than seen in observations. 

\subsection{Observational effects} 
To make a realistic comparison between the results for the SAM and the SDSS, different observational effects must be taken into account. We thus build mock catalogs from the SAM data following \citet{li07}. We first test the influence of fiber collisions 
which reduce the number of redshift measurements for galaxies in close pairs by about $6\%$. Because our smoothing scale
is nearly always larger than $55''$, fiber collisions do not affect our analysis significantly. 

The most notable influences from obervations are the redshift distortion effect and the effect from the survey geometry. 
The redshift-space distortion affects mainly the high density regions and plays a role of ``mixing''. 
In redshift space, galaxies in massive clusters present a Finger-of-God
configuration along the line of sight. On the other hand, galaxies in filaments surrounding a massive cluster have
a tendency to be flowing into the cluster, which squashes the filaments along the line of sight.
In other words, galaxies in filaments and galaxies in massive clusters are somewhat mixed in redshift space. Thus
the color scatter in a given density bin becomes larger in redshift space, resulting in a lower color-density correlation
coefficient in high density regions, as shown by the green solid line in Figure 8. 
Furthermore, the redshift-space distortion makes the measured environmental density systematically lower than that measured
in real space. The ellipticity calculated in the redshift space is systematically lower/higher for regions with 
high/low $e$ in real space. That is, filaments become rounder and clusters tend to be more anisotropic in redshift space.
Such effects cause color mixing in the $[e,\ln(1+\delta)]$ plane, and thus lead to flatter color contours with respect to the 
ellipticity $e$ and weaker correlations with $\delta$ (see the black contour lines in Figure 9).

In low density regions, measurements of environmental density and ellipticity are mainly affected by 
survey geometry. Both are biased to lower values. 
In the left panel of Figure 10, we show the value of $e$ with respect to the distance to the 
survey boundary for SDSS sample $M_r$20. Different lines represent different environmental density ranges with 
longer dashes corresponding to higher densities. Physically, 
we do not expect any correlations between $e$ value and the distance to the survey boundary. 
For high-density lines, they are indeed nearly flat. For low-density lines, however, 
the measured $e$ is systematically lower for galaxies closer to the survey boundary.
The effect is significant for galaxies that are within $20h^{-1}\hbox{ Mpc}$ of the boundary. 
This is clearly an observational effect due to padding the regions outside the survey boundary with the 
average number density. Such a measurement bias on $e$ can lead to artificial correlations between 
galaxy properties and environmental ellipticity. The significance of the effect depends on the
the fraction of galaxies that are within $20h^{-1}\hbox{ Mpc}$ of the boundary. 
Volume-limited samples with smaller volume are affected more. In the right panel of 
Figure 10, we show the ratio of color in high $e$ regions to the color in low $e$ regions vs. 
minus the absolute magnitude of galaxies -$M_r$. Here high and low $e$ are defined as above the $1\sigma$ level 
and below the $-1\sigma$ level with respect to the average $e$ (see the right panel of Figure 1). 
We present results for different SDSS volume-limited samples, $M_r$18, $M_r$19 and $M_r$20 from smaller to larger 
volume. For $M_r$18, we see the decrease of the color ratio with respect to -$M_r$ of galaxies. 
That is, for relatively faint galaxies, the ones in low $e$ regions tend to be bluer than those in 
high $e$ regions. For bright galaxies, the ratio is close to $1$. For $M_r$19 and $M_r$20, 
there is no detectable dependence of the color ratio on -$M_r$,
and the ratio is always close to $1$. To test if the $M_r$-dependence of the color ratio 
seen in $M_r$18 is physical or due to the boundary effect, we construct
subsamples of galaxies with distances less than $110h^{-1}\hbox{ Mpc}$ from $M_r$19 and $M_r$20, respectively. 
This cut corresponds roughly to the distance limit of $M_r$18. The results for the $M_r$19 and $M_r$20 subsamples 
are shown by unfilled green triangles and unfilled red squares, respectively. The apparent correlation
with $M_r$ of galaxies is then seen. This demonstrates that the correlation seen for $M_r$18 sample
is largely due to the boundary effect. As shown in the left panel of Figure 10, in low density regions, 
the $e$ value is biased to a lower value for galaxies closer to the survey boundary. Because  
galaxies in low density regions tend to be bluer than those in high density regions, this bias leads to
an increase in the number of relatively blue galaxies in low $e$ regions, and thus increases the color ratio
between high $e$ and low $e$ regions. The increase of the color ratio is more significant for fainter galaxies
because the observed ones reside in a smaller volume around the observer and thus suffer more from the boundary effect. 

In Figure 11, we show color contours in the $[e_o, \ln(1+\delta_o)]$ plane for $M_r$18.
For $M_r$18, the optimal adaptive smoothing scale for a galaxy corresponds to the distance to its $8$th nearest neighbor.
Comparing to the middle panel of Figure 4, we see a notable correlation between color and ellipticity
in low-density regions. This however, is largely due to the survey boundary effect that biases both
the environmental density and ellipticity toward lower values in low density regions.

\begin{figure}
\includegraphics[width=42mm]{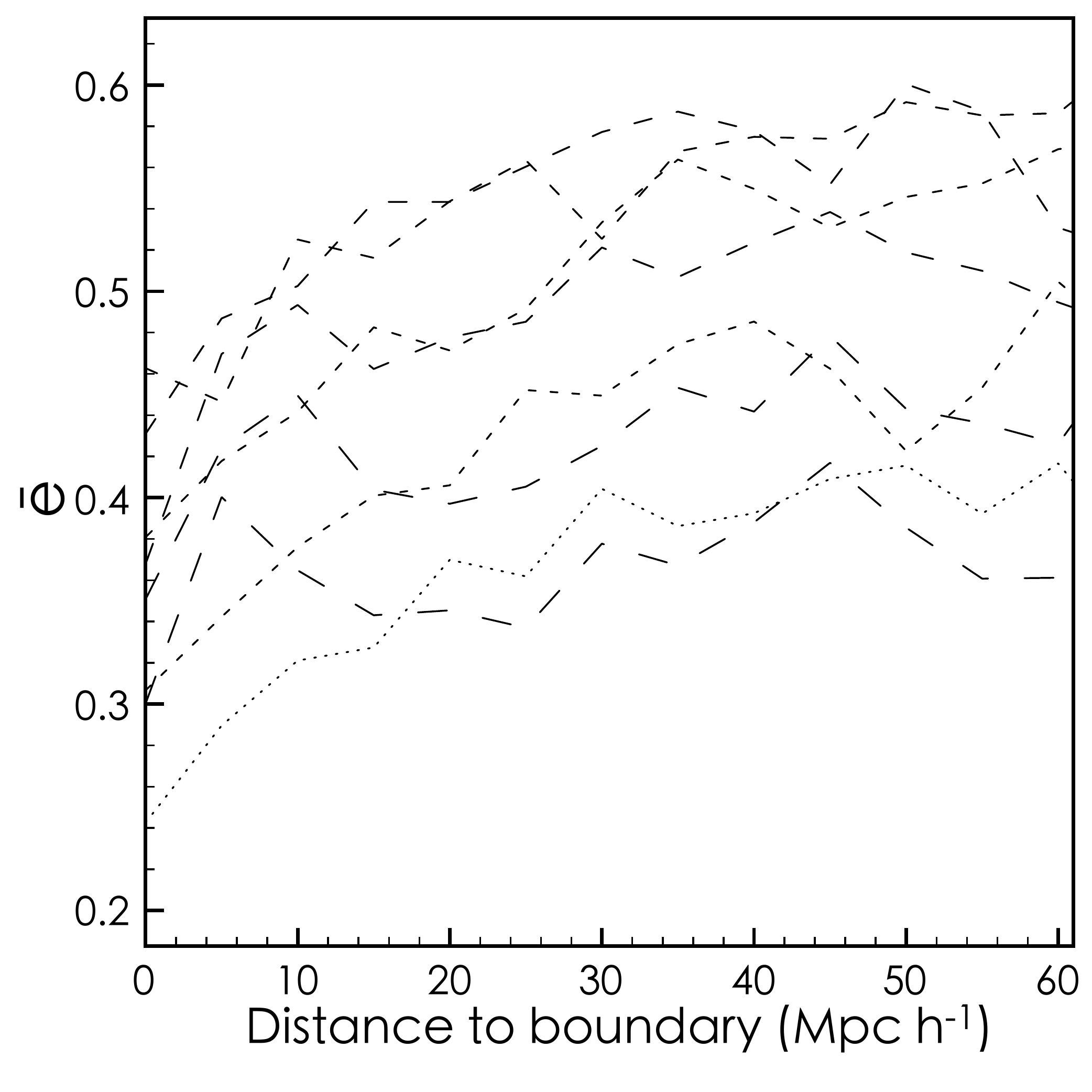}\includegraphics[width=42mm]{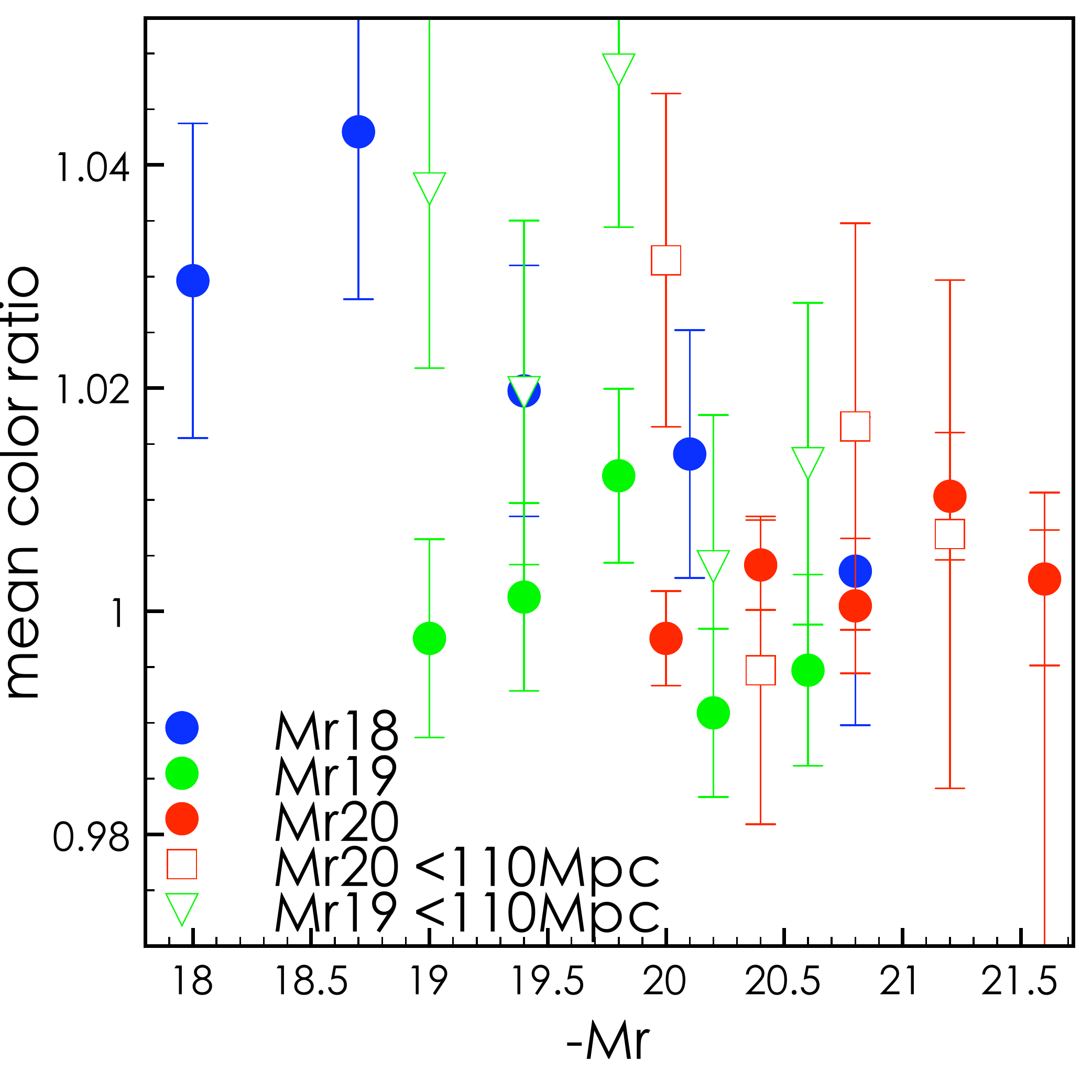}
 \caption{Left panel: The average environment ellipticity for galaxies as a function of their distance to 
the survey boundary. The optimal adaptive smoothing has been applied. Different lines represent the results
for different bins of $log(1+\delta_o)$ ranging from $\log(1+\delta_o)=-2.0$ to $\ln(1+\delta_o)=4.0$. Longer-dashed lines 
represent the results for higher density bins. Right panel: The ratio of the mean $(g-r)$ in high ellipticity 
regions to that in low ellipticity regions for $M_r$ 20 (red filled circle), $M_r$19 (green filled circle) and $M_r$18 samples 
(blue filled circle), respectively.  The high and low ellipticity regions are defined as the regions above the $+1\sigma$ dashed line
and below the $-1 \sigma$ dashed line shown in the right pane of Figure 1. 
The results for subsamples of $M_r$20 and of $M_r$19 cut at $110h^{-1}\hbox{ Mpc}$ are shown by red open squares and green open triangles, respectively.
}
 \end{figure}

\begin{figure}
\includegraphics[width=84mm]{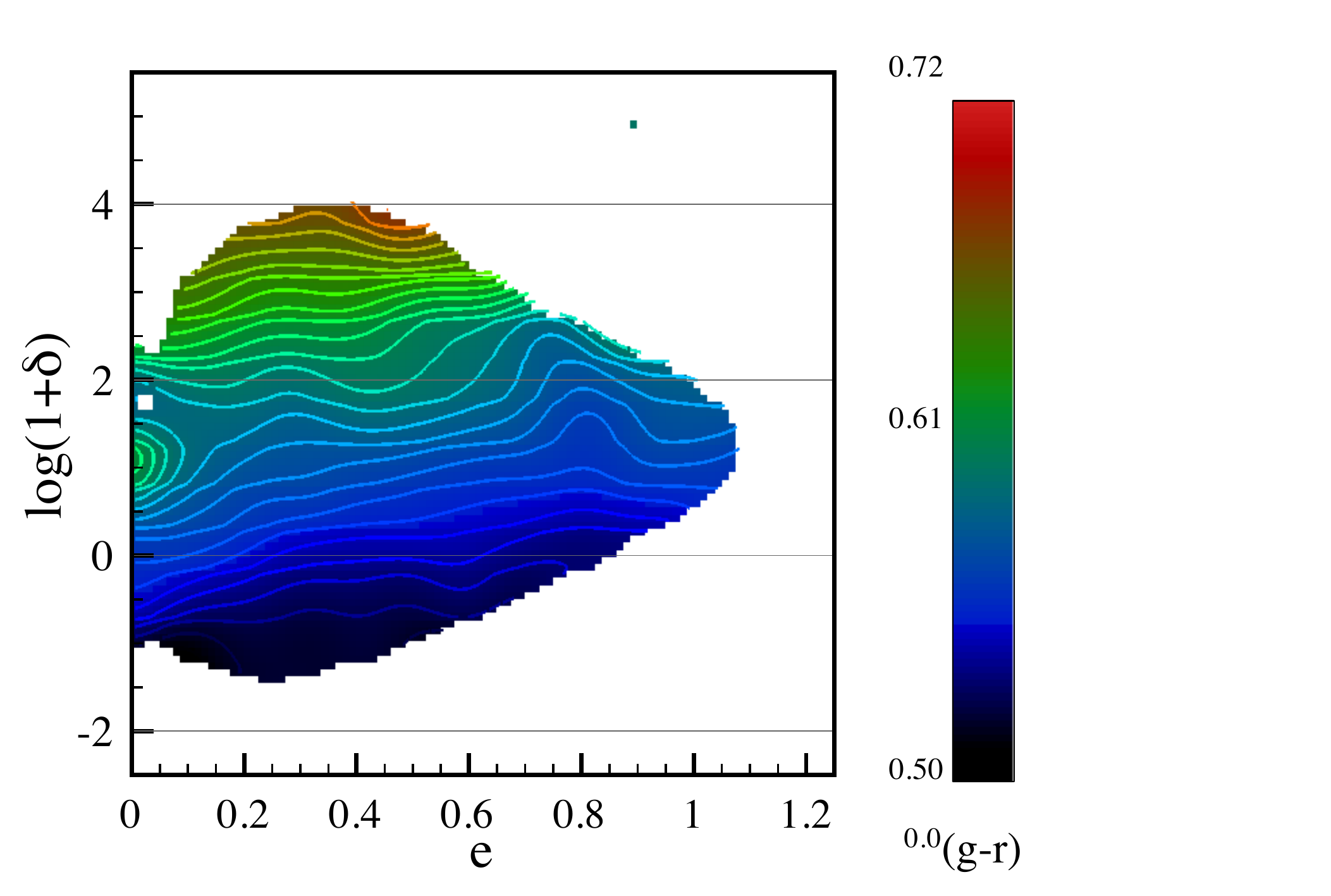}
 \caption{Galaxy color contours for SDSS $M_r$18. The environmental smoothing scale for a galaxy is taken to be the 
distance to its $8th$-nearest neighbor. The boundary is the $0.3\sigma$ uncertainty contour, where $\sigma$ = 0.168 is the standard deviation of 
galaxy color of the whole $M_r$18 sample.
}
 \end{figure}

\section{Conclusion and discussion}

In this paper we examine the tidal dependence of galaxy properties in the NYU-VAGC sample and compare 
it with the predictions of the SAM. To separate the environment-morphology dependence
from the environmental-density dependence, we construct the density fluctuation $\delta$, the ellipticity $e$ 
and the prolateness $p$ from the three eigenvalues of the tidal tensor
of the potential field calculated from the spatial distribution of galaxies, where $e$ and $p$ are nearly
independent of $\delta$. It should be noted that this potential field can be biased with respect to the 
potential field from the underlying dark matter distribution. The well-known galaxy bias refers to the ratio of the
density fluctuation amplitude of galaxy distribution and that of the dark matter distribution.

For $M_r$20, the volume-limited sample with absolute magnitude brighter than $-20$, 
our results show that while the environmental-density dependence of galaxy properties is indisputable,  
the dependence on $e$ and $p$ is rather weak, except perhaps for galaxy size. 
It is further shown that such weak correlations depend
on the smoothing scale considered for the environment. There exists a particular scale, 
$\sim 2 h^{-1}\hbox{ Mpc}$, at which, the environmental-$e$ ($p$) dependence nearly vanishes. This corresponds to the optimal adaptive smoothing scale
 where the environmental-density dependence of galaxy properties is the strongest. It varies from high density regions to
low density regions, and corresponds well to the distance to each galaxy's $3rd$-nearest-neighbor for $M_r20$ sample.  
At smoothing scales larger/smaller than this `optimal smoothing scale', the weak correlations between galaxy properties and $e$
are negative/positive. We demonstrate, however, that these correlations result mainly from correlations between $(e,\delta)$ at the
considered smoothing scale and the environmental density $\delta_o$ at the optimal adaptive smoothing scale. In other words, 
we find no physical influence of environment morphology on galaxy properties. Furthermore, it is $\delta_o$ that 
plays the dominant role for the environmental effects, and the apparent density dependence on other smoothing scales 
is largely due to the correlations between $\delta$ and $\delta_o$. This indicates that, galaxy properties in $M_r$20
are affected mainly by their nearby environments. Our analysis shows that 
galaxy size is independent of environment in high-density regions. In low-density regions, some
correlation with $e_o$ in addition to the dependence on $\delta_o$ is detected.
Galaxies in lower $(\delta_o, e_o)$ regions tend to be smaller. Such correlations need to be further
explored with future observations.

With NYU-VAGC $M_r$19, \citet{blanton07} analyze galaxies in groups, and find that galaxy properties depend mainly on the properties of
their host groups, and are not affected independently by the environment on larger scales($> 1h^{-1}\hbox{ Mpc}$).
As they only consider galaxies in groups, it is appropriate to compare their results with ours in high density regions.
From Figure 8, we can see that the `peak scale' in high density regions is $\sim 1h^{-1}\hbox{ Mpc}$, consistent
with the typical group scale in their studies. Therefore our results are in agreement
with theirs concerning the dominant importance of the nearby environment. This in turn
validates the current halo occupation distribution model, which assumes that galaxy properties depend
only on those of their parent halos with no additional influence from larger scale.  
\citet{lee08} also examines the tidal dependence of galaxy properties. They found that there is a tendency
for elliptical galaxies to be preferentially situated in low $e$ regions with density in the range $0.5\le \delta\le 1.06$.
For $-0.3\le \delta\le 0.1$, spiral galaxies are likely in regions with large $e$. We note that their considered scale
for environmental effects is $\sim 400/64 \sim 6 h^{-1}\hbox{ Mpc}$. Thus their results can loosely be
compared with our analysis for $R_s=3h^{-1}\hbox { Mpc}$ (but note the different $\delta$ range). With the 
correspondence between color and morphology of galaxies, the positive correlation
of galaxy color with $e$ seen in right panel of Figure 4 is qualitatively consistent with the trend found by \citet{lee08}
although their definition of $e$ is different from ours. On the other hand, our analysis indicates that
this $e$-dependence is largely due to the correlations between $e$ and $\delta_o$ and therefore has 
little physical significance. 

Comparing the SDSS results with those from our SAM, we see qualitative agreement between the two,
although the SAM predicts a stronger dependence of galaxy properties on environmental density.
Our results are also in accordance with the theoretical study of \citet{desjacques08}, which 
showed that environmental effects on dark matter halo formation mainly reflect environmental density, 
and are influenced little by $e$ and $p$. \citet{hahn09} found that the anti-correlation between
the formation epoch of galactic halos and their environment density is, in fact, mainly attributable to tidal suppression by 
neighboring massive halos. Because of the enhanced environment density near massive halos, this suppression shows up as an assembly bias, and provides 
a possible explanation for galaxy variation with environment density. On the other hand, \citet{hahn09} chose
the smallest eigenvalue $\lambda_3$ of the tidal tensor as an indicator of the strength of tidal field, and they 
found a stronger dependence of halo formation epoch on $\lambda_3$ than on environment density. Our results
manifest a strong degeneracy between $\lambda_3$  and environment density. Just as for
ellipticity, $\lambda_3$ can be geometrically more strongly correlated with environment density on the optimal 
smoothing scale than on a larger scale. Further work is needed to clarify whether the dependence
of halo formation history on environment morphology can be attributed to this geometry effect, or whether baryon physics actually make difference
between halos and galaxies. 

By constructing mock catalogs from SAM data, we have examined the influence of observational effects on 
our analysis. Redshift-space distortion is significant in high density regions. It mixes galaxies
in filaments and galaxies in clusters causing an increase of color scatter at given density,
and therefore reduces the color-density correlation coefficient considerably. 
The survey boundary affects environment measurements mainly in low density regions, biasing 
both the environmental density and the ellipticity to lower values. The effect is 
notable for galaxies within $20h^{-1}\hbox{ Mpc}$ of the survey boundary.
While it is negligible for $M_r$20, the effect is considerable for $M_r$18 which occupies a relatively small
volume and therefore has a large fraction of its galaxies close to the boundary. It induces an
artificial correlation between galaxy properties and the environmental ellipticity for $M_r$18.  

In summary, our analysis of SDSS data shows that in addition to environment density, there is no significant further dependence of galaxy properties 
on the morphology of large scale structure. Geometrically, both ellipticity and density on one smoothing scale 
correlate strongly with each other, and with ellipticity/density on other smoothing scales. If the smoothing scale is not chosen 
properly, an appeared dependence of galaxy properties on both environment ellipticity and environment density arises which is merely due to geometry. 
We find 
that for the optimal adaptive smoothing scale, the dependence on density is maximized and the dependence on ellipticity and prolateness is null.

\section*{Acknowledgments}

We thank Cristiano Porciani and Guinevere Kauffmann for detailed discussions in the early stages of 
this project, and Cheng Li for his help on dealing with the SDSS data. This research is supported in part by the NSFC of China under grants
10773001 and 11033005, 11173001 and the 973 program No. 2007CB815401, as well as by the ERC
under Advanced Grant 246797 ``Galformod''.

\bsp

\label{lastpage}

\appendix
\section[]{Scale dependence of color-density relation}

Previous studies have shown that age-related properties, such as galaxy color, 
are correlated with environmental density and that this correlation is scale dependent
\citep{kauffmann04,blanton07,park07}. Here we
examine the scale dependence of the color-density relation quantitatively and in more detail. 

For each galaxy, we calculate its environmental density at two different smoothing scales,
denoted as $\delta_{R_1}$ and $\delta_{R_2}$. We then analyze the dependence
of galaxy color on $(\delta_{R_1}, \delta_{R_2})$. The results are shown in Figure A1, 
where $R_1=2h^{-1}\hbox{ Mpc}$ is taken to be the reference scale.
In each panel, the black dots show the values of $[\ln(1+\delta_{R_1}), \ln(1+\delta_{R_2})]$ for all the galaxies,
and the red lines are $(g-r)$ contours. The green line denotes our dividing line for high and 
low density regions, which is the $-45^o$ line passing through the point of 
$(1.0,d_0)$, where $d_0$ is the corresponding 
value of $\ln(1+\delta_{R_2})$ given $\ln (1+\delta_{2 Mpch^{-1}})=1.0$.   
It is seen that for $R_2\ge 5h^{-1}\hbox{ Mpc}$,
the color-density dependence, $(g-r)-(\delta_{2 Mpch^{-1}},\delta_{R_2})$,
is mainly on $\delta_{2 Mpch^{-1}}$, and the additional dependence on $\delta_{R_2}$ is very weak.
For $R_2= 3h^{-1}\hbox{ Mpc}$, a certain level of $\delta_{R_2}$-dependence is seen in low density regions.
For $R_2= 1h^{-1}\hbox{ Mpc}$, on the other hand, the $\delta_{R_2}$-dependence appears in high
density regions. These results clearly demonstrate the scale dependence of the color-density relation.
There exists a special scale where the environmental density affects galaxy colors the most.
The value of this `optimal adaptive smoothing scale' is $\sim 2h^{-1}\hbox{ Mpc}$ for $M_r$20,  
and varies somewhat from high to low density regions. Our detailed analysis show
that the variation is from $\sim 1.5h^{-1}\hbox{ Mpc}$ in high density regions to 
$\sim 2.5h^{-1}\hbox{ Mpc}$ in low density regions, consistent with that shown in Figure 8
for the scale dependence of the color-density correlation coefficient. 

Our further studies find that a better way to quantify the environmental effects
on a galaxy is to use adaptive smoothing, for example distance to its n$th$-nearest neighbor, which takes into account the 
variations of the environmental density naturally. We have tested the distributions for the 
n$th$-nearest neighbor distances, where $n=1, 2, 3, 4$ are considered. 
We notice that the distance to the $3rd$-nearest neighbor for $M_r20$ sample
is peaked at $2 h^{-1}\hbox{ Mpc}$ with a range mainly between $\sim 1 h^{-1}\hbox{ Mpc}$ and
$3h^{-1}\hbox{ Mpc}$, therefore it represents the `peak scale' suitably and adaptively, and we propose 
it as the optimal adaptive smoothing scale for the $M_r20$ sample. The corresponding optimal adaptive smoothing scales 
for the $M_r19$ and $M_r18$ samples are respectively 6th-nearest and 8th-nearest neighbors. 

\begin{figure}
\includegraphics[width=84mm]{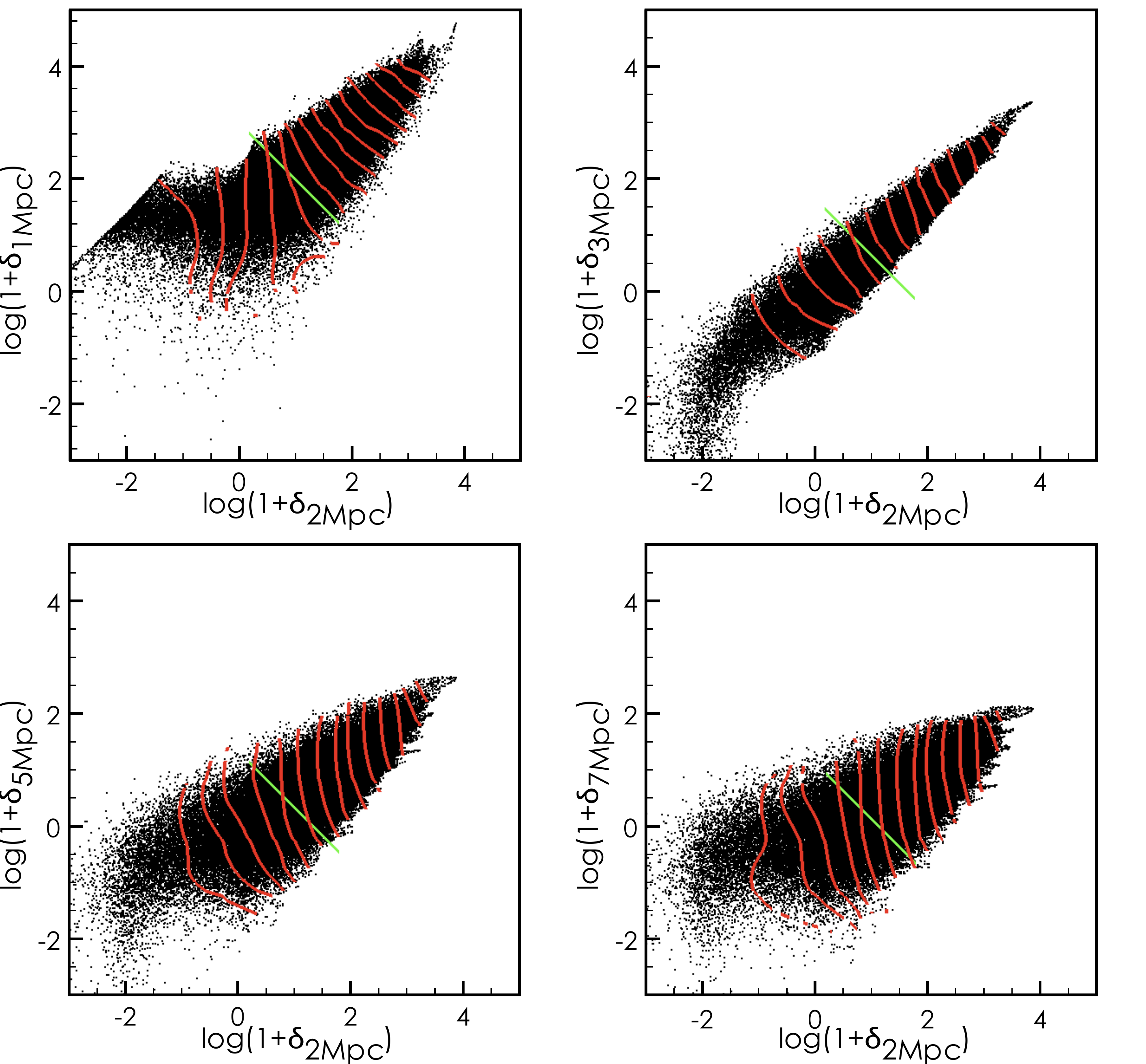}
 \caption{Scatter plots of $[\ln(1+\delta_{R_1}), \ln(1+\delta_{R_1})]$.
In each panel, the red lines are galaxy color contours and the green line denotes the dividing line
for high and low density regions.}
\end{figure}
\end{document}